\documentclass[12pt]{article}

\usepackage{multirow}
\usepackage{amsmath,color}
\usepackage{commath}
\usepackage{amsthm}
\usepackage{amssymb}
\usepackage{setspace,mathrsfs}
\usepackage{algorithm,natbib}
\usepackage{bm}
\usepackage{placeins}

\usepackage{array}
\usepackage{ragged2e}
\usepackage{booktabs}
\usepackage{overpic}

\newcolumntype{P}[1]{>{\RaggedRight\arraybackslash}p{#1}}

\usepackage{url,amsmath,amsfonts,amssymb}
\usepackage{caption,subcaption,algorithm,graphicx}
\usepackage{listings,enumerate,color,relsize,multirow,rotating,algorithm}
\usepackage{booktabs}
\usepackage{bbm}
\usepackage{algpseudocode}
\usepackage{adjustbox}
\usepackage{tabularx}
\usepackage{alphalph}
\usepackage{comment}
\usepackage[normalem]{ulem}
\usepackage[
  colorlinks=true,
  linkcolor=blue,
  filecolor=blue,
  citecolor=blue,
  urlcolor=blue
]{hyperref}

\let\emptyset\varnothing

\newtheorem{assume}{Assumption}
\usepackage{mathrsfs}

\def \R {\mathbb{R}}

\newtheorem{proposition}{Proposition}

\newtheorem{coro}{Corollary}
\newtheorem{theorem}{Theorem}

\begin{document}

\def\spacingset#1{\renewcommand{\baselinestretch}%
{#1}\small\normalsize} \spacingset{1}

{
  \title{\bf Graph-Based Change-Point Detection for Partially Observed High-Dimensional Data}
  \author{Mingshuo Liu and Hao Chen\thanks{Corresponding author. Email: \texttt{hxchen@ucdavis.edu}.}\\
  Department of Statistics, University of California, Davis}
  \date{}
  \maketitle
}

\bigskip

\begin{abstract}
Partial missingness is common in high-dimensional data, but most existing change-point procedures are developed for fully observed sequences. We introduce \textit{gMiss}, a graph-based framework for testing and localizing a change in the observed-data distribution of a partially observed high-dimensional sequence. The method treats the observed values together with the missingness indicators as the object of inference, so the target alternative is a change in the induced observed data law. It is designed for general distributional changes and requires neither sparsity nor Gaussianity. When the augmented observations are independent, the full permutation test controls type I error in finite samples. The procedure combines graph scans based on elementwise imputation and distance imputation. The two scans capture complementary graph patterns. Simulation results indicate that \textit{gMiss} maintains accurate null calibration across the MCAR and MAR designs considered, remains competitive under Gaussian location alternatives, and exhibits strong power and localization performance in many non-Gaussian location and scale settings. We further illustrate the practical utility of the method through an application to genomic copy-number data, where \textit{gMiss} identifies additional candidate boundaries that are visually plausible in the raw heatmap.
\end{abstract}

\noindent%

\vfill

\newpage

\spacingset{1.8}

\section{Introduction}
\label{sec:intro}

Change-point detection (CPD) seeks to determine whether an ordered sequence of
observations is homogeneous and, if not, to estimate where its distribution changes.
The ordering may represent time, genomic position, spatial location, or another
meaningful index, so CPD arises in diverse settings such as high-throughput neural
recordings \citep{chen2019universal,zhang2021graph}, climate and environmental
monitoring \citep{reeves2007review,shi2022changepoint}, network monitoring
\citep{peel2015detecting,barnett2016change,wang2021optimal}, and financial modeling
\citep{andreou2009structural,allen2018non,banerjee2020change}.

A broad class of modern CPD methods is nonparametric, aiming to avoid strong
distributional assumptions that may be unrealistic in complex data. Representative
examples include kernel-based methods
\citep{harchaoui2007retrospective,harchaoui2008kernel,li2015m,garreau2018consistent,arlot2019kernel,song2024practical},
interpoint distance-based methods \citep{matteson2014nonparametric,li2020asymptotic},
graph-based methods \citep{chen2015graph,chu2019asymptotic,zhang2021graph,liu2022fast},
methods based on self-normalization
\citep{wang2022inference,zhang2022adaptive,jiang2023robust}, and ranking-based
methods \citep{zhou2025asymptotic}. Despite this progress, most high-dimensional CPD
procedures are designed for fully observed sequences.

In many modern applications, however, the observations are both high-dimensional and
only partially observed. Sensor networks may exhibit partial dropouts due to failures
or communication issues \citep{loh2011high}, gene expression matrices often contain
missing values due to experimental limitations \citep{troyanskaya2001missing}, and
electronic health record data frequently contain substantial missing information due
to the way data are recorded in routine care \citep{wells2013strategies}. {
Existing approaches to CPD with missingness address important but distinct settings. For
example, \citet{xie2012change} studied streaming data that lie near a changing
{low-dimensional} manifold and {detected} changes through tracking residuals, while
\citet{londschien2021change} developed {likelihood-based} procedures for changes in
Gaussian graphical models with missing values. These works are related in that they
handle missing observations, but their targets are different from the partially observed
{high-dimensional} vector sequences considered here.

Closer to our setting, \citet{follain2022high} proposed the \textit{MissInspect} framework,
which uses a projected CUSUM construction to estimate sparse {high-dimensional} mean changes
under heterogeneous MCAR missingness. However, it does not provide a finite-sample $p$-value.
\citet{liu2025high} proposed a four-step procedure that combines blockwise
imputation, candidate screening, and local refinement to estimate the number and locations
of multiple sparse mean changes under MCAR missingness. The method is not formulated as
a global hypothesis test with a $p$-value for the observed data null.

Concretely, we target general changes in the observed data law without
sparsity or Gaussianity assumptions, while also providing a change-point estimate and
finite-sample-valid permutation calibration.
}

In some straightforward approaches, missing entries are handled by simple heuristics originally designed for complete data, such as dropping coordinates with high missing rates or imputing by coordinatewise means or local averages.
While such steps can be \mbox{convenient}, they can reduce power or lead to invalid calibration because imputation may weaken the change-point signal or induce dependence among observations. The difficulty is not imputation itself, but whether the calibration reflects the imputation step. If imputation depends on the ordering of the observations and is performed only once, a later complete data calibration does not reproduce the same dependence. Permutation validity can still hold if permuting before or after imputation gives the same result, or if imputation and all subsequent steps are repeated for every permuted dataset. To illustrate these issues, we consider the following experiment.

Consider $n=200$ observations, where each observation is a 50-dimensional vector with i.i.d.\ coordinates.
We generate a missing data version by independently masking each entry under {an MCAR mechanism with} missing rate $0.3$.
We consider two benchmark procedures designed for fully observed data: the max-type edge-count test (\textit{MET}) \citep{chen2019change} and the E-Divisive method (\textit{ecp}) \citep{matteson2014nonparametric}. For comparability, we apply coordinatewise mean imputation to the missing entries
before running these benchmark methods: for each coordinate, every missing entry is
replaced by the average of the observed entries in that coordinate. We evaluate empirical size under a heavy-tailed null with i.i.d.\ $t_2$ coordinates, and empirical power under (i) a location alternative, where coordinates change from $t_2$ to $t_2+0.35$ after the midpoint, and (ii) a scale alternative, where coordinates change from $t_2$ to $1.18\times t_2$ after the midpoint.
The experiment is repeated 400 times, and the results for the benchmark methods are summarized in Table~\ref{tab:moti}.

\begin{table}[!htbp]
\caption{Rejection rates under the null and the location and scale alternatives ($\alpha = 0.05$) for complete and missing data. For the missing data setting, we apply coordinatewise mean imputation before running the complete data benchmark methods.}
\label{tab:moti}
\centering
\begin{tabular}{c|c|cc}
\midrule \midrule
\textbf{Scenario} & \textbf{Data} & \textit{MET} & \textit{ecp} \\
\midrule
\multirow{2}{*}{null}
& complete & 0.0650 & 0.0450 \\
& missing  & 0.0550 & 0.0550 \\
\midrule
\multirow{2}{*}{\shortstack{location\\alternative}}
& complete & 0.8375 & 0.4600 \\
& missing  & 0.5650 & 0.2400 \\
\midrule
\multirow{2}{*}{\shortstack{scale\\alternative}}
& complete & 0.7500 & 0.3575 \\
& missing  & 0.5900 & 0.2900 \\
\midrule \midrule
\end{tabular}
\end{table}

Table~\ref{tab:moti} suggests that both benchmark methods achieve reasonable empirical size in this setting.
However, applying coordinatewise mean imputation followed by the original complete data procedure can reduce power, especially under the location alternative.
The reduction is also evident under the scale alternative.

A second common heuristic is local neighbor averaging (e.g., averaging the $m$ closest observations in index within each coordinate).
If this imputation is performed once and then a complete data procedure is applied with its usual calibration, empirical sizes can become severely inflated.
To illustrate this, we impute each missing entry by averaging the nearest $m\in\{2,4,6\}$ observed neighbors in the same coordinate and then apply standard complete data CPD methods.
The resulting empirical sizes are summarized in Table~\ref{tab:moti2}. They are far above the nominal level when the missing rate is $0.3$, and remain inflated even when the missing rate is reduced to $0.1$.

\begin{table}[!htbp]
\caption{Empirical sizes at nominal level $\alpha = 0.05$ for \textit{MET} and \textit{ecp},
under two missing rates and different numbers of neighbors $m$ used in the local neighbor averaging imputation.
All results are computed under the null setting with no change-point.}
\label{tab:moti2}
\centering
\begin{tabular}{c|cc|cc}
\midrule \midrule
\multirow{2}{*}{$\bm{m}$ }
& \multicolumn{2}{c|}{\textbf{missing rate = 0.3}}
& \multicolumn{2}{c}{\textbf{missing rate = 0.1}} \\
\cline{2-5}
& \textit{MET} & \textit{ecp}
& \textit{MET} & \textit{ecp} \\
\midrule
2 & 0.9875 & 0.9825 & 0.2925 & 0.4575 \\
4 & 0.9975 & 0.9875 & 0.3050 & 0.3875 \\
6 & 1.0000 & 0.9975 & 0.3075 & 0.3750 \\
\midrule \midrule
\end{tabular}
\end{table}

Together, Tables~\ref{tab:moti} and \ref{tab:moti2} show that missing value treatment and calibration must be designed together. In Table~\ref{tab:moti}, coordinatewise mean imputation leaves empirical size near the nominal level but reduces power. Table~\ref{tab:moti2} shows a different problem: when local imputation is performed once and then held fixed during complete data calibration, empirical type I error is severely inflated. These findings motivate methods that either repeat all data-dependent steps under permutation or avoid imputation that depends on the ordering.

Graph-based CPD provides a natural starting point. The weighted and difference edge-count scans capture complementary edge-count patterns, with the former particularly sensitive to location changes and the latter to scale changes \citep{chu2019asymptotic}. We retain both components but adapt graph construction and permutation calibration to missing observations.

\section{Data model, missingness, and hypotheses}
\label{sec:data-setting}

\noindent \textbf{Complete and observed data hypotheses.} Let $\mathbf{X}=[X_{it}] \in \mathbb{R}^{p\times n}$ denote the (unobserved) complete data matrix,
where $X_t\in\mathbb{R}^p$ denotes the observation at position $t$ in the sequence and $[n]=\{1,\ldots,n\}$. We assume throughout that the complete data observations are independent across sequence positions.
A common formulation of the single change-point problem is to test homogeneity of the sequence,
\begin{equation}
H_0:\ X_t \sim F_0,\quad t\in[n],
\label{eqn:null-complete}
\end{equation}
against the alternative
\begin{equation}
H_1:\ \exists\,\tau\in[n-1]\ \text{such that}\
X_t\sim F_0\ \text{for }t\leq\tau,\quad
X_t\sim F_1\ \text{for }t>\tau,
\label{eqn:alt-complete}
\end{equation}
where $F_0$ and $F_1$ are distinct distributions on $\mathbb{R}^p$.

To model missing values, we introduce an indicator matrix $\mathbf{M}=[M_{it}] \in \{0,1\}^{p\times n}$,
where $M_{it}=1$ if $X_{it}$ is observed and $M_{it}=0$ if it is missing. Define the partially
observed vector at position $t$ as
$X_t^{\mathrm{obs}} = X_t \circ M_t,
$
where $\circ$ denotes the {Hadamard product}. Thus, the observed information at position
$t$ is the pair $(X_t^{\mathrm{obs}}, M_t)$. For notational convenience, we encode this pair as an
augmented vector
\begin{equation}
X_t^\ast = \bigl((X_t^{\mathrm{obs}})^\top,\ M_t^\top\bigr)^\top \in \mathbb{R}^{2p}.
\label{eq:xtstar-def}
\end{equation}
Including $M_t$ in $X_t^\ast$ ensures that zeros in $X_t^{\mathrm{obs}}$ are not confounded with
missing entries.

The missing data mechanism can be described through the conditional law of $\mathbf{M}$
given $\mathbf{X}$. For reference, under missing completely at random (MCAR), the
missingness pattern is independent of the complete data, whereas under missing at random
(MAR), it may depend on the observed part of the data but not on its unobserved part.
Our validity results are not tied to a particular MCAR or MAR model;
instead, they require independence of the augmented observations.

\begin{assume}[Independence]
\label{ass:independence}
The augmented observations $X_1^\ast,\ldots,X_n^\ast$ are independent.
\end{assume}

We conduct testing and calibration on the observed augmented sequence
$\{X_t^\ast\}_{t=1}^n$. Accordingly, we formulate the observed data hypotheses
\begin{equation}
H_0:\ X_t^\ast \sim F_0^\ast,\quad t\in[n],
\label{eqn:nullmissing}
\end{equation}
against the single change-point alternative
\begin{equation}
H_1:\ \exists\,\tau\in[n-1]\ \text{such that}\
X_t^\ast\sim F_0^\ast\ \text{for }t\le\tau,\quad
X_t^\ast\sim F_1^\ast\ \text{for }t>\tau,
\label{eqn:altmissing}
\end{equation}
where $F_0^\ast$ and $F_1^\ast$ denote the induced distributions of $X_t^\ast$ under the pre- and
post-change regimes, respectively, with $F_0^\ast\neq F_1^\ast$. { These observed data hypotheses are not equivalent to the complete data hypotheses in
\eqref{eqn:null-complete}--\eqref{eqn:alt-complete}. Since the augmented vector includes both observed values and missingness indicators, the alternative may arise from a change in the observed values, a change in the observation pattern, or both. Conversely, a change in the complete data distribution may be weakened or hidden after missingness is applied. In the numerical studies below, we impose location and scale changes on the complete data distribution before missingness is applied. We do not systematically study alternatives driven mainly by changes in the missingness mechanism.}

Under $H_0$ in \eqref{eqn:nullmissing}, Assumption~\ref{ass:independence}
makes the augmented observations i.i.d.\ and hence exchangeable, as required for full
permutation validity. The specific MCAR and MAR designs used below satisfy the assumption
because, conditional on the complete data, their missingness indicators are independent
across positions.

\section{Graph-based detection framework for missing data}
\label{Section: framework}
{We first review graph-based change-point methods for complete data, and then present the construction used by \textit{gMiss} to handle missingness.}

\subsection{A concise review of graph-based change-point approaches for complete data}
\label{subsec:review-complete}

Graph-based change-point detection for complete data was introduced by
\citet{chen2015graph} and extended to more general settings by
\citet{chu2019asymptotic}. Under the complete data null in
\eqref{eqn:null-complete}, the observations are i.i.d.\ and hence exchangeable, so
the inference can be based on the permutation null distribution. We write
$\mathbf{E}$ and $\mathbf{Var}$ for expectation and variance under this distribution.

Given a distance on the observations, one constructs a similarity graph $G$ on
$\{X_t\}_{t=1}^n$, such as a $k$-minimum spanning tree ($k$-MST), a
$k$-nearest neighbor graph, or a minimum distance pairing graph \citep{rosenbaum2005exact}. For a candidate
split point $t$, define
$
R_1(t)=\sum_{(i,j)\in G}\mathbf{1}\{i\le t,\ j\le t\},
R_2(t)=\sum_{(i,j)\in G}\mathbf{1}\{i>t,\ j>t\}
$
and
$
R_0(t)=\sum_{(i,j)\in G}
\mathbf{1}\{i\le t<j\ \text{or}\ j\le t<i\}.
$
Here $R_1(t)$ and $R_2(t)$ count within-segment edges, while $R_0(t)$ counts
between-segment edges. Then $R_0(t)+R_1(t)+R_2(t)=|G|$, where $|G|$ represents the number of edges in $G$.

The original graph-based scan statistic of \citet{chen2015graph} is based on the
standardized between-segment edge count
\[
Z_0(t)
=
\frac{\mathbf{E}\{R_0(t)\}-R_0(t)}
{\sqrt{\mathbf{Var}\{R_0(t)\}}}.
\]
A large value of $Z_0(t)$ indicates that fewer graph edges cross the split than
expected under the permutation null, which is typical when the two segments are
well separated.

The original scan can suffer from variance boosting when a candidate split
produces segments of unequal sizes. In moderate or high dimensions, it can also
lose power under scale changes because the between-segment edge count need not
be small. To address these two limitations, \citet{chu2019asymptotic}
considered the following weighted and difference components:
\begin{align}
\label{eq:Zw,def}
Z_w(t)
&=
\frac{R_w(t)-\mathbf{E}\{R_w(t)\}}
{\sqrt{\mathbf{Var}\{R_w(t)\}}},
\qquad
{R_w(t)=\omega_1(t)R_1(t)+\omega_2(t)R_2(t),}
\end{align}
where
{$\omega_1(t)=\frac{n-t-1}{n-2}$ and $\omega_2(t)=\frac{t-1}{n-2}$},
and
\begin{align}
\label{eq:Zd,def}
Z_d(t)
&=
\frac{R_d(t)-\mathbf{E}\{R_d(t)\}}
{\sqrt{\mathbf{Var}\{R_d(t)\}}},
\qquad
R_d(t)=R_1(t)-R_2(t).
\end{align}
The weights in $R_w(t)$ account for unequal segment sizes. Under a location
change, both within-segment counts tend to exceed their permutation-null
expectations, so $Z_w(t)$ tends to be large. Under a scale change, the segment
with smaller spread tends to have an excess of within-segment edges, whereas
the segment with larger spread tends to have a deficit, so $|Z_d(t)|$ tends to
be large.
Since the type of change is usually unknown in practice, using both components
can protect against relying on a single edge-count pattern.
This motivates the max-type scan statistic $M(t)=\max\{Z_w(t), |Z_d(t)|\}$,
which is the complete data scan statistic \citep{chu2019asymptotic} that our missing data procedures adapt.
\subsection{Ingredients for the proposed framework}
\subsubsection{Local elementwise imputation and valid {type I error} control}
\label{Subsubsection: treatment 1}

Because the change-point location is unknown, imputation schemes that average broadly across
the sequence may blur the contrast between pre- and post-change regimes. Our first route, termed
elementwise imputation (\textit{EI}), therefore imputes each missing entry using only nearby
observed values from the same coordinate.

{
Let $m\in\mathbb{N}$ be an even integer chosen by the user. For a missing entry with $M_{it}=0$,
let $\mathcal{N}_i(t)\subseteq\{s\in[n]:M_{is}=1\}$ denote the selected neighboring sequence positions
in coordinate $i$: we choose up to $m$ indices, with at most $m/2$ closest observed indices
from each side of $t$. Thus, $|\mathcal{N}_i(t)|$ may be smaller than $m$ near the boundaries
or when coordinate $i$ is sparsely observed. For each $s\in\mathcal{N}_i(t)$, define
\[
w_{it}(s)
=
\frac{\exp(-|s-t|/2)}
{\sum_{r\in\mathcal{N}_i(t)}\exp(-|r-t|/2)}.
\]
The imputed value is $\tilde X_{it}=\sum_{s\in\mathcal{N}_i(t)}w_{it}(s)X^{\mathrm{obs}}_{is}$.
}
We use exponentially decaying weights so that nearby observations receive more weight and distant
observations are less likely to blur a potential change-point. Other weight choices can also be considered. Let $\widetilde{\mathbf X}=(\tilde X_{it})\in\mathbb R^{p\times n}$ denote
the imputed matrix in Algorithm~\ref{Algorithm: elementwise imputation Method}.

\begin{algorithm}[!htbp]
\renewcommand{\thealgorithm}{1}
\caption{Elementwise Imputation (EI)}
\label{Algorithm: elementwise imputation Method}
\footnotesize
\textbf{Input:} Partially observed data $\mathbf{X}^\ast=(\mathbf{X}^{\mathrm{obs}},\mathbf{M})$, even integer $m$ \\
\textbf{Output:} Imputed matrix $\widetilde{\mathbf{X}}$
\begin{algorithmic}[1]
\State Initialize $\widetilde{\mathbf{X}} \gets \mathbf{X}^{\mathrm{obs}}$.
\For{each coordinate $i \in \{1,\dots,p\}$}
  \If{$\sum_{t=1}^n M_{it}=0$}
      \State Set $\tilde X_{i1},\ldots,\tilde X_{in}\gets 0$
      \State \textbf{continue}
  \EndIf
  \For{each sequence position $t \in \{1,\dots,n\}$ with $M_{it}=0$}
    \State Count observed values before/after $t$:
    $a_- := \sum_{k=1}^{t-1} M_{ik}$,\quad $a_+ := \sum_{k=t+1}^{n} M_{ik}$.
    \State Choose neighbor counts:
    {$b_- := \min(\frac{m}{2}, a_-)$,\quad $b_+ := \min(\frac{m}{2}, a_+)$.}
\State {Let $\mathcal{N}_i(t)$ contain the $b_-$ nearest observed positions before $t$ and $b_+$ nearest after $t$.}
\State {For $s\in\mathcal{N}_i(t)$, set $w_{it}(s)\propto\exp(-|s-t|/2)$ and normalize $\sum_{s\in\mathcal{N}_i(t)}w_{it}(s)=1$.}
\State {Impute $\tilde X_{it} \gets \sum_{s\in\mathcal{N}_i(t)} w_{it}(s)X^{\mathrm{obs}}_{is}$.}
  \EndFor
\EndFor
\State \Return $\widetilde{\mathbf{X}}$.
\end{algorithmic}
\end{algorithm}

The neighborhood size $m$ controls the locality of the elementwise imputation. In all numerical experiments and in the real data analysis below, we use the default value $m=6$, corresponding to at most three observed neighbors before and three observed neighbors after the target position. Although \textit{EI} is simple, it does not fit directly into asymptotic results of the complete data analytic framework. First, multiple missing entries may be imputed from overlapping local
neighborhoods, inducing dependence within each coordinate. Second, the similarity graph used
by graph-based CPD itself depends on the data: after permuting the sequence and recomputing
\textit{EI}, the resulting graph is generally not a fixed relabeling of the original graph.
Therefore, the analytic permutation null mean and variance formulas derived for fixed graphs
constructed once from complete data are not directly applicable. Reusing them can lead to
invalid calibration, as illustrated in Table~\ref{tab:moti2}. We therefore calibrate the scan statistic based on \textit{EI} through a permutation scheme that recomputes the imputation and graph construction for each permuted ordering.

\begin{algorithm}[!htbp]
\renewcommand{\thealgorithm}{2}
\caption{{Elementwise Imputation Permutation Framework (EI-Perm)}}
\label{algo:PEI}
\footnotesize
\textbf{Input:} Partially observed data $\mathbf{X}^\ast$, nonempty scan range $[n_0,n_1]\subseteq\{1,\ldots,n-1\}$, neighborhood size $m$, number of permutations $B$, raw graph-based scan score $Q(t)$ \\
\textbf{Output:} approximate Monte Carlo permutation $p$-value $\hat p$, change-point estimate $\hat\tau$
\begin{algorithmic}[1]
\State Compute $\widetilde{\mathbf{X}} \gets \mathrm{EI}(\mathbf{X}^\ast,m)$ and construct the graph $G$ based on EI.
\State Compute the observed raw scan score $Q_{\mathrm{obs}}(t)$ for all $n_0\le t\le n_1$ from $(\widetilde{\mathbf{X}},G)$.
\For{$b=1,\ldots,B$}
    \State Draw a random permutation $\pi_b$ and form $\mathbf{X}^\ast_{\pi_b}$ by permuting columns of $\mathbf{X}^\ast$.
    \State Compute $\widetilde{\mathbf{X}}_{\pi_b} \gets \mathrm{EI}(\mathbf{X}^\ast_{\pi_b},m)$ and construct the corresponding graph $G^{(b)}$.
    \State Compute the permuted raw scan score $Q^{(b)}(t)$ for all $n_0\le t\le n_1$ from $(\widetilde{\mathbf{X}}_{\pi_b},G^{(b)})$.
\EndFor
\State Compute
$$
\hat\mu_Q(t)=\frac{1}{B}\sum_{b=1}^B Q^{(b)}(t), \quad \hat\sigma_Q^2(t)=\frac{1}{B-1}\sum_{b=1}^B\bigl(Q^{(b)}(t)-\hat\mu_Q(t)\bigr)^2
$$
for all $n_0\le t\le n_1$.
\State Form the standardized scan score
$$
Z_{\mathrm{obs}}(t)=\frac{Q_{\mathrm{obs}}(t)-\hat\mu_Q(t)}{\hat\sigma_Q(t)}, \quad
Z^{(b)}(t)=\frac{Q^{(b)}(t)-\hat\mu_Q(t)}{\hat\sigma_Q(t)}.
$$
\State {Define the score to be maximized by}
{
$$
A_{\mathrm{obs}}(t)=
\begin{cases}
Z_{\mathrm{obs}}(t), & Q(t)=R_w(t),\\
|Z_{\mathrm{obs}}(t)|, & Q(t)=R_d(t),
\end{cases}
\qquad
A^{(b)}(t)=
\begin{cases}
Z^{(b)}(t), & Q(t)=R_w(t),\\
|Z^{(b)}(t)|, & Q(t)=R_d(t).
\end{cases}
$$
}
\State Set
$$
{
T_{\mathrm{obs}}=\max_{n_0\le t\le n_1} A_{\mathrm{obs}}(t), \quad
T^{(b)}=\max_{n_0\le t\le n_1} A^{(b)}(t), \quad
\hat\tau=\arg\max_{n_0\le t\le n_1} A_{\mathrm{obs}}(t).
}
$$
\State Compute the approximate Monte Carlo permutation $p$-value
$
\hat p = \frac{1}{B}\sum_{b=1}^B \mathbf{1}\{T^{(b)}\geq T_{\mathrm{obs}}\}.
$
\end{algorithmic}
\end{algorithm}

Let $\mathbf{X}^\ast_{\pi} = (X_{\pi(1)}^\ast,\dots,X_{\pi(n)}^\ast)$ denote the matrix obtained by permuting the columns of
$\mathbf{X}^\ast$ according to a permutation $\pi$ of $\{1,\ldots,n\}$. Fix a scan range
$[n_0,n_1]$ and the number of Monte Carlo permutations $B$. For the observed data and for
each permuted dataset, we recompute \textit{EI}, rebuild the similarity graph, and evaluate
the same raw graph-based scan score over the resulting ordered sequence. Because the graph
changes across permutations under \textit{EI}, the permutation mean and variance of the raw
scan score are estimated empirically at each $t$ and then used for standardization. The final
test statistic is obtained by scanning the standardized score over $t\in[n_0,n_1]$.

Algorithm~\ref{algo:PEI} summarizes the full \textit{EI-Perm} framework. When $Q(t)=R_w(t)$,
this yields the weighted scan. When $Q(t)=R_d(t)$, it yields the {two-sided difference scan based on $|Z_d(t)|$}. For the
max-type scan, the same idea is applied componentwise to $R_w(t)$ and $R_d(t)$: we first
standardize them separately using their permutation sample means and variances, and then
combine them through $\max\{Z_w(t),|Z_d(t)|\}$. The exact statement for finite samples proved in
Section~\ref{sec:theory} concerns the corresponding full permutation test, while Algorithm~\ref{algo:PEI}
uses a Monte Carlo approximation to that exact procedure.

\subsubsection{Pairwise distance imputation}
\label{Subsubsection: treatment 2}

{\textit{DI} imputes pairwise distances in a permutation-equivariant way: permuting the sequence indices only permutes the rows and columns of the imputed distance matrix. Under the stagewise uniqueness condition stated below, the graph is constructed once and permutation replicates are generated by relabeling its vertices.}

Let $\mathbf{D}=[D_{st}]_{s,t=1}^n$ denote the unobserved complete distance matrix, where
$D_{st}$ is the Euclidean distance between $X_s$ and $X_t$. We construct an imputed
distance matrix $\widetilde{\mathbf{D}}=[\widetilde D_{st}]_{s,t=1}^n$ from the observed pair
$\mathbf{X}^\ast=(\mathbf{X}^{\mathrm{obs}},\mathbf{M})$. For each pair $(s,t)$, let
$
C_{st}=\sum_{i=1}^p M_{is}M_{it}
$
be the number of jointly observed coordinates. When $C_{st}>0$, we estimate the squared
distance using the overlap coordinates and rescale by $p/C_{st}$:
\begin{equation}
\label{eq:DI_def}
\widetilde D_{st}
=
\Big(
\frac{p}{C_{st}}
\sum_{i=1}^p
M_{is}M_{it}
\bigl(X^{\mathrm{obs}}_{is}-X^{\mathrm{obs}}_{it}\bigr)^2
\Big)^{1/2}.
\end{equation}
The factor $p/C_{st}$ rescales the squared distance computed from the jointly observed coordinates to the full dimension. Under homogeneous MCAR, the jointly observed coordinates form a random subset of coordinates, so this rescaling gives a natural estimate of the full Euclidean distance when coordinates are on comparable scales. Under heterogeneous MCAR or more general missingness mechanisms, the same approximation is most appropriate when the jointly observed coordinates remain approximately representative of all coordinates for the pairwise squared differences. {This representativeness condition concerns the quality of the distance approximation, whereas the permutation validity results below rely on Assumption~\ref{ass:independence} together with the observed data null.}

\begin{algorithm}[!htbp]
\caption{Pairwise Distance Imputation (\textit{DI})}
\label{Algorithm: Pairwise Distance Imputation}
\footnotesize
\begin{algorithmic}[1]
\Require Partially observed data $\mathbf{X}^{\ast}=(\mathbf{X}^{\mathrm{obs}},\mathbf{M})$
\Ensure Imputed distance matrix $\widetilde{\mathbf D}$

\For{all $1\le s<t\le n$}
    \State $C_{st}\gets \sum_{i=1}^p M_{is}M_{it}$.
    \If{$C_{st}>0$}
        \State $\widetilde D_{st} \gets \Big(\frac{p}{C_{st}} \sum_{i=1}^p M_{is}M_{it}(X^{\mathrm{obs}}_{is}-X^{\mathrm{obs}}_{it})^2\Big)^{1/2}$.
    \EndIf
\EndFor

\State $\mathcal I_+\gets\{(s,t):1\le s<t\le n,\ C_{st}>0\}$.

\If{$\mathcal I_+=\emptyset$}
    \State Set $\widetilde D_{st}\gets 1$ for all $1\le s<t\le n$.
\Else
    \State $D_{\mathrm{med}}\gets
    \operatorname{median}\{\widetilde D_{st}:(s,t)\in\mathcal I_+\}$.
    \For{all $1\le s<t\le n$ with $C_{st}=0$}
        \State $\widetilde D_{st}\gets D_{\mathrm{med}}$.
    \EndFor
\EndIf

\State Set $\widetilde D_{ss}\gets 0$ and symmetrize by setting $\widetilde D_{ts}\gets \widetilde D_{st}$ for all $1\le s<t\le n$.
\State \Return $\widetilde{\mathbf D}$.
\end{algorithmic}
\end{algorithm}

When $C_{st}=0$, no direct information is available, so we impute $\widetilde D_{st}$ by a
global summary, namely the median of the available imputed distances. If no pair has positive
overlap, all off-diagonal distances are set to one.
Algorithm~\ref{Algorithm: Pairwise Distance Imputation} summarizes pairwise distance
imputation (\textit{DI}). We then construct a similarity graph $G$ (e.g., a $k$-MST) from
$\widetilde{\mathbf{D}}$. The next proposition formalizes the permutation-equivariance property described above.

\begin{proposition}
\label{prop:DI-equivariant}
Let $\widetilde{\mathbf D} = \textit{DI}(\mathbf{X}^\ast)$ and
$\widetilde{\mathbf D}^\pi = \textit{DI}(\mathbf{X}^\ast_{\pi})$ denote the \textit{DI}
distance matrices computed from $\mathbf{X}^\ast$ and $\mathbf{X}^\ast_{\pi}$,
respectively. Let $P_\pi$ be the permutation matrix defined by
$P_\pi e_s=e_{\pi(s)}$, where $e_1,\ldots,e_n$ are the standard basis vectors in
$\mathbb R^n$. Then, for all $s,t \in [n]$,
$
\widetilde D^\pi_{st} = \widetilde D_{\pi(s),\pi(t)},
\text{equivalently}~
\widetilde{\mathbf D}^\pi = P_\pi^\top\, \widetilde{\mathbf D}\, P_\pi.
$
\end{proposition}

{For the finite-sample validity result for the fixed-graph \textit{DI} route, we assume that every successive MST used to construct the graph exists and is unique. Under this condition, Proposition~\ref{prop:DI-equivariant} implies that the graph associated with any permuted ordering is exactly a relabeling of the original graph. The \textit{DI} graph can therefore be constructed once and reused throughout permutation calibration.}

{In contrast to \textit{EI-Perm}, the \textit{DI} route does not require estimating permutation means and variances empirically. Because the graph is fixed once $\widetilde{\mathbf D}$ is constructed, the chosen scan score admits analytic standardization, so the corresponding complete data formulas remain available in our setting.} In particular, letting $G_i$ denote the set of
edges incident to vertex $i$ and writing $|G|$ for the total number of edges, the
permutation null mean and variance of $R_w(t)$ and $R_d(t)=R_1(t)-R_2(t)$ are
\begin{align*}
& \mathbf{E}\big(R_w(t)\big) = |G| \frac{(t-1)(n-t-1)}{(n-1)(n-2)}, \qquad
\mathbf{E}\big(R_d(t)\big) = |G| \frac{2 t - n}{n},\\
& \mathbf{Var}\big(R_w(t)\big) =
\frac{t(t-1)(n-t)(n-t-1)}{n(n-1)(n-2)(n-3)}
\Big(|G| - \frac{\sum_{i=1}^n |G_i|^2}{n-2} + \frac{2|G|^2}{(n-1)(n-2)} \Big),\\
& \mathbf{Var}\big(R_d(t)\big) =
\frac{t(n-t)}{n(n-1)}
\Big(\sum_{i=1}^n |G_i|^2 - \frac{4|G|^2}{n}\Big).
\end{align*}
These yield the standardized weighted and difference scan scores introduced earlier, and the
max-type score is then obtained by combining them. More generally, {the \textit{DI-Perm} route}
can be coupled with different graph-based scan scores. {Algorithm~\ref{algo:PPDI} summarizes the {\textit{DI-Perm}} framework. Neither distance imputation nor graph construction is repeated within the permutation loop.}

\begin{algorithm}[!htbp]
\renewcommand{\thealgorithm}{4}
\caption{Distance Imputation Permutation Framework (DI-Perm)}
\label{algo:PPDI}
\footnotesize
\textbf{Input:} Partially observed data $\mathbf{X}^\ast$, nonempty scan range $[n_0,n_1]\subseteq\{1,\ldots,n-1\}$, number of permutations $B\ge1$, {standardized scan score $S(t)$, with $S(t)=Z_w(t)$ for the weighted scan, $S(t)=|Z_d(t)|$ for the difference scan, and $S(t)=M(t)$ for the max-type scan} \\
\textbf{Output:} approximate Monte Carlo permutation $p$-value $\hat p$, change-point estimate $\hat\tau$
\begin{algorithmic}[1]
\State Apply DI to $\mathbf{X}^\ast$ to obtain $\widetilde{\mathbf D}$, {construct the fixed graph $G$}, and compute the observed maximum scan statistic
$T_{\mathrm{obs}}=\max_{n_0\le t\le n_1} S_{\mathrm{obs}}(t),
\quad
\hat\tau=\arg\max_{n_0\le t\le n_1} S_{\mathrm{obs}}(t).
$
\For{$b=1,\ldots,B$}
    \State {Draw a permutation $\pi_b$, relabel the vertices of the fixed graph $G$ according to $\pi_b$, and compute the corresponding maximum scan statistic}
$   T^{(b)}=\max_{n_0\le t\le n_1} S^{(b)}(t).$
\EndFor
\State Compute the approximate Monte Carlo permutation $p$-value
$\hat p=\frac{1}{B}\sum_{b=1}^B \mathbf{1}\{T^{(b)}\ge T_{\mathrm{obs}}\}$.
\end{algorithmic}
\end{algorithm}

\begin{table}[!htbp]
\centering
\caption{Rejection rates of \textit{EI-Perm} and \textit{DI-Perm} equipped with {weighted} ($Z_w(t)$), {difference} ($|Z_d(t)|$), and max-type ($M(t)$) scans under location and scale alternatives ($\alpha=0.05$).}
\label{tab:EWIPDIcomp}
\begin{tabular}{c|cc|c|cc|c}
\midrule
\midrule
\multirow{2}{*}{Alternative} & \multicolumn{3}{c|}{\textit{EI-Perm}} & \multicolumn{3}{c}{\textit{DI-Perm}} \\
\cline{2-7}
 & $Z_w(t)$ & {$|Z_d(t)|$} & $M(t)$ & $Z_w(t)$ & {$|Z_d(t)|$} & $M(t)$ \\
\midrule
location alternative & 0.7325 & 0.0550 & 0.6900 & 0.5575 & 0.0450 & 0.4550 \\
scale alternative & 0.0950 & 0.7025 & 0.5225 & 0.2550 & 0.8950 & 0.8000 \\
\midrule
\midrule
\end{tabular}
\end{table}

Revisiting the motivating settings in Section~\ref{sec:intro}, the results in Table~\ref{tab:EWIPDIcomp} show complementary behavior across the two imputation routes and scan scores. Under the location alternative, \textit{EI-Perm} combined with the {weighted scan statistic} $Z_w(t)$ has the highest power, while under the scale alternative, {\textit{DI-Perm} combined with the difference scan statistic $|Z_d(t)|$} has the highest power. The max-type statistic $M(t)$ provides a compromise across the two types of alternatives, though it is not uniformly best in either setting. These findings motivate the combined procedure in Section~\ref{sec:combined,approach}.

\FloatBarrier

\subsubsection{Optional robust preprocessing (robust standardization)}
\label{subsubsec:rs}

When coordinate scales differ substantially, distance-based graphs can be dominated by a
small subset of high-variance coordinates, reducing sensitivity to changes in other
coordinates. To mitigate this, we apply an optional robust standardization step prior to
imputation. The goal is to rescale each coordinate using a local variability estimate that
is minimally affected by a single change-point.

Let $\{t_1<\cdots<t_{e_i}\}$ denote the observed sequence positions in coordinate $i$
(i.e., $M_{i t_k}=1$). When $e_i\ge 2$, define
$$
s_i^2 = \frac{1}{2(e_i-1)}\sum_{k=1}^{e_i-1}
\bigl(X^{\mathrm{obs}}_{i t_k}-X^{\mathrm{obs}}_{i t_{k+1}}\bigr)^2,
\quad s_i=\sqrt{s_i^2},
$$
and set $s_i=1$ when $e_i<2$ or when the displayed formula gives $s_i=0$. Algorithm~\ref{algo: robustsd} summarizes the standardization procedure.

We use this standardization step as an optional fixed preprocessing step to stabilize
coordinate scales before imputation and graph construction. The main simulation results
in Section~\ref{Section: simulation} and the main real data analysis in
Section~\ref{Section: real data} are reported without this step. Its effect is examined
in Supplementary Material, and the robustly standardized version of the real
data analysis is also reported there. When robust standardization
is used, the scale factors are computed once from the observed ordering and the
permutation procedure is then applied to the standardized augmented sequence. Because these
scale factors depend on the observed ordering, this fixed RS variant is an empirical
preprocessing option and is not covered by Corollary~\ref{cor:gmiss-permutation}, which covers the version without RS, or a version that recomputes RS within every ordering.

\begin{algorithm}[!htbp]
\renewcommand{\thealgorithm}{5}
\caption{Robust Standardization (RS)}
\label{algo: robustsd}
\footnotesize
\textbf{Input:} Partially observed data $\mathbf{X}^\ast= (\mathbf{X}^{\mathrm{obs}},\mathbf{M})$ \\
\textbf{Output:} Standardized partially observed data $\mathbf{X}^\ast_{\mathrm{sd}}$
\begin{algorithmic}[1]
\State Initialize $\mathbf{X}^{\mathrm{obs}}_{\mathrm{sd}} \gets \mathbf{X}^{\mathrm{obs}}$.
\For{each coordinate $i \in \{1,\dots,p\}$}
    \State Let $t_1<\cdots<t_{e_i}$ be the indices with $M_{i t_k}=1$.
    \If{$e_i \ge 2$}
        \State $s_i^2 \gets \frac{1}{2(e_i-1)} \sum_{k=1}^{e_i-1}\big(X^{\mathrm{obs}}_{i t_k}-X^{\mathrm{obs}}_{i t_{k+1}}\big)^2$; set $s_i \gets \sqrt{s_i^2}$.
        \If{$s_i=0$}
            \State $s_i \gets 1$.
        \EndIf
    \Else
        \State $s_i \gets 1$.
    \EndIf
    \For{each observed index $t \in \{t_1,\ldots,t_{e_i}\}$}
        \State $X^{\mathrm{obs}}_{\mathrm{sd},it} \gets X^{\mathrm{obs}}_{it}/s_i$.
    \EndFor
\EndFor
\State Output $\mathbf{X}^\ast_{\mathrm{sd}}$ as the observed pair $(\mathbf{X}^{\mathrm{obs}}_{\mathrm{sd}},\mathbf{M})$ (equivalently, augment as in \eqref{eq:xtstar-def}).
\end{algorithmic}
\end{algorithm}

\subsection{Proposed approaches}
\label{subsec:proposed}
Throughout, we use the $k$-MST as the similarity graph. The $k$-MST is the union of
$k$ minimum spanning trees constructed successively on the same vertex set, with edges
selected in earlier trees excluded when computing each subsequent tree.

The tuning parameter $k$ controls the graph density, and the performance of edge-count scan statistics can be sensitive to this choice. Early work in multivariate graph tests and
graph-based CPD typically recommends a small constant (e.g., $k=5$) as a default
choice \citep{friedman1979multivariate, chu2019asymptotic}. More recently,
\citet{zhu2024limiting} studied denser graph regimes and showed that allowing
$k = O(n^\eta)$ for $0<\eta<1$ can yield substantial power gains in a range of
settings, with empirical evidence suggesting $\eta \approx 0.5$ as a robust choice.
In the missing data setting, distances are estimated with additional noise and effective
information can be reduced. Motivated by this, we adopt a slightly denser graph by taking
$k=\lfloor n^{0.55}\rfloor$, which is also supported by numerical experiments in Section~\ref{sec:choiceofk}.

\subsubsection{A combined approach}
\label{sec:combined,approach}

{We now combine the elementwise and pairwise distance imputation routes to obtain our main procedure, \textit{gMiss}. Robust standardization may be applied beforehand when coordinate scales differ substantially.} The two components in \textit{gMiss} use different graph constructions for missing data. The weighted component is obtained from the \textit{EI} route: after elementwise imputation and graph construction, we compute the raw weighted edge-count statistic $R_w^{\mathrm{EI}}(t)$ and standardize it empirically using its permutation replicates to obtain $Z_w^{\mathrm{EI}}(t)$. The difference component is obtained from the \textit{DI} route: after pairwise distance imputation and graph construction, we use the analytic permutation formulas for the fixed graph to compute the standardized difference statistic $Z_d^{\mathrm{DI}}(t)$. Thus, the combined scan statistic is $M(t)=\max\{Z_w^{\mathrm{EI}}(t),\, |Z_d^{\mathrm{DI}}(t)|\}$.
The first component is typically sensitive to location changes, whereas the second component is typically sensitive to scale changes in moderate to high dimensions. Algorithm~\ref{algo:combined} summarizes the procedure.

\begin{algorithm}[!htbp]
\renewcommand{\thealgorithm}{6}
\caption{Bi-Imputation Permutation Framework (gMiss)}
\label{algo:combined}
\footnotesize
\textbf{Input:} Partially observed data $\mathbf{X}^\ast$, nonempty scan range $[n_0,n_1]\subseteq\{1,\ldots,n-1\}$, neighborhood size $m$, number of permutations $B$ \\
\textbf{Output:} approximate Monte Carlo permutation $p$-value $\hat p$, change-point estimate $\hat\tau$
\begin{algorithmic}[1]
\State (Optional) Compute $\mathbf{X}^\ast_{\mathrm{sd}} \gets \mathrm{RS}(\mathbf{X}^\ast)$; if skipped, set $\mathbf{X}^\ast_{\mathrm{sd}} \gets \mathbf{X}^\ast$.
\State Compute the raw weighted edge-count statistic $R_w^{\mathrm{EI}}(t)$ from \textit{EI} on $\mathbf{X}^\ast_{\mathrm{sd}}$ for all $n_0\le t\le n_1$.
\State {Construct the fixed \textit{DI} graph from $\mathbf{X}^\ast_{\mathrm{sd}}$ and compute the analytically standardized difference statistic $Z_d^{\mathrm{DI}}(t)$ for all $n_0\le t\le n_1$.}
\For{$b=1,\ldots,B$}
    \State Draw a random permutation $\pi_b$ and form $\mathbf{X}^\ast_{\mathrm{sd},\pi_b}$ by permuting columns of $\mathbf{X}^\ast_{\mathrm{sd}}$.
    \State Compute $R_w^{\mathrm{EI},(b)}(t)$ for all $n_0\le t\le n_1$ from \textit{EI} on $\mathbf{X}^\ast_{\mathrm{sd},\pi_b}$.
    \State {Compute $Z_d^{\mathrm{DI},(b)}(t)$ for all $n_0\le t\le n_1$ by relabeling the vertices of the fixed \textit{DI} graph according to $\pi_b$.}
\EndFor
\State Compute sample mean $\hat\mu_w^{\mathrm{EI}}(t)$ and sample standard deviation $\hat\sigma_w^{\mathrm{EI}}(t)$ from $\{R_w^{\mathrm{EI},(b)}(t)\}_{b=1}^B$.
\State Form
\[
Z_w^{\mathrm{EI}}(t)=\frac{R_w^{\mathrm{EI}}(t)-\hat\mu_w^{\mathrm{EI}}(t)}{\hat\sigma_w^{\mathrm{EI}}(t)},\qquad
Z_w^{\mathrm{EI},(b)}(t)=\frac{R_w^{\mathrm{EI},(b)}(t)-\hat\mu_w^{\mathrm{EI}}(t)}{\hat\sigma_w^{\mathrm{EI}}(t)}.
\]
\State Form
\[
M(t)=\max\{Z_w^{\mathrm{EI}}(t),\,|Z_d^{\mathrm{DI}}(t)|\},\qquad
M^{(b)}(t)=\max\{Z_w^{\mathrm{EI},(b)}(t),\,|Z_d^{\mathrm{DI},(b)}(t)|\}
\]
for all $n_0\le t\le n_1$ {and $b=1,\ldots,B$}.
\State Set $T_M=\max_{n_0\le t\le n_1} M(t)$ and {$T_M^{(b)}=\max_{n_0\le t\le n_1} M^{(b)}(t)$ for $b=1,\ldots,B$}.
\State Change-point estimate: $\hat\tau=\arg\max_{n_0\le t\le n_1} M(t)$.
\State Approximate Monte Carlo permutation $p$-value:
$\hat p=\frac{1}{B}\sum_{b=1}^B \mathbf{1}\{T_M^{(b)}\ge T_M\}$.
\end{algorithmic}
\end{algorithm}

\subsubsection{A fast approach}
\label{sec:fast-app}

While \textit{gMiss} has good power in several settings below, it can be computationally intensive because \textit{EI} and the resulting graph must be recomputed within each permutation. To reduce runtime, we propose a fast alternative, \textit{gMiss-F}, which relies only on \textit{DI} and uses the analytic tail approximation for the max-type edge-count scan developed by \citet{chu2019asymptotic} to obtain $p$-values without Monte Carlo permutations. We treat \textit{gMiss-F} as a computational approximation. In this paper, its calibration is assessed empirically through size experiments rather than established by a theorem covering finite samples. The detailed null settings and empirical sizes of \textit{gMiss-F} are reported in Supplementary Material, and the rejection probabilities remain close to the nominal level across the regimes considered there.

The comparison in Table~\ref{tab:slowfastcomp} uses the motivating setting in Section~\ref{sec:intro}. Table~\ref{tab:slowfastcomp} places the combined procedures \textit{gMiss} and \textit{gMiss-F} alongside separate max-type procedures for the \textit{EI} and \textit{DI} routes. Under the location alternative, the max procedure based on \textit{EI} has higher power than the corresponding \textit{DI} procedure, and \textit{gMiss} matches the performance of \textit{EI} in this example. Under the scale alternative, the max procedure based on \textit{DI} has the highest power, and \textit{gMiss-F} remains close to it, while \textit{gMiss} provides an intermediate performance. Overall, these results are consistent with the complementary roles of the two routes: \textit{EI} is more effective for location changes, whereas \textit{DI} is more effective for scale changes in this experiment. In practice, \textit{gMiss} is the default choice when computation permits and the change signature is unknown, whereas \textit{gMiss-F} is a faster approximation based on \textit{DI} whose calibration should be checked empirically for the application at hand. We use \textit{gMiss} in the numerical studies and real data illustration reported in Sections~\ref{Section: simulation}~and~\ref{Section: real data}.

\begin{table}[!htbp]
\centering
\caption{Rejection rates of the max-type procedures for the two routes and the combined procedures under location and scale alternatives ($\alpha = 0.05$). Here \textit{EI-Perm} and \textit{DI-Perm} denote the max-type scan $M(t)$ applied within the \textit{EI} and \textit{DI} routes, respectively.}
\label{tab:slowfastcomp}
\begin{tabular}{c|cccc}
\midrule
\midrule
Alternative & \textit{EI-Perm}  & \textit{DI-Perm}  & \textit{gMiss} & \textit{gMiss-F} \\
\midrule
location alternative & 0.6900 & 0.4550 & 0.6900 & 0.4450 \\
scale alternative & 0.5225 & 0.8000 & 0.7375 & 0.7950 \\
\midrule
\midrule
\end{tabular}
\end{table}

To provide some intuition for these complementary behaviors, recall the two canonical change patterns for graph-based edge-count scans. Under a location change, observations are typically more similar within each segment than across segments near the true change-point, which leads to both within-segment edge counts
$R_1(t)$ and $R_2(t)$ exceeding their permutation null expectations when $t$ is near the true change-point $\tau$. Consequently, the weighted edge-count scan
$Z_w(t)$ tends to take large positive values near the change.
\textit{EI} is well matched to this regime because it imputes each missing entry using only
temporally nearby observations within the same coordinate by a locally weighted average, reducing the extent to which observations from different segments are mixed during imputation. This local reconstruction better preserves the neighborhood structure within segments that drives the regular pattern signal captured by $Z_w(t)$.

In contrast, under a scale change, especially in moderate to high dimensions, the similarity graph can exhibit a different configuration: observations from the segment with higher variance may connect to neighbors in the segment with lower variance, creating an imbalance between $R_1(t)$ and $R_2(t)$ when $t$ is near the true change-point $\tau$, so that $|Z_d(t)|$ can be large while $Z_w(t)$ is not.
\textit{DI} is well matched to this regime because it estimates interpoint distances from overlap coordinates and rescales by $p/C_{st}$. When the overlap coordinates are representative, this can make an increase in scale visible at the distance level and can produce a graph geometry that amplifies the $R_1(t)$ versus $R_2(t)$ imbalance targeted by the difference scan $Z_d(t)$. In contrast, \textit{EI}'s local averaging can partially attenuate variance inflation by smoothing extreme values. These considerations help explain why the \textit{EI} component is useful for location shifts, whereas a \textit{DI} approach can be particularly competitive under scale alternatives.

\section{Theoretical results}
\label{sec:theory}

We begin by establishing a general permutation validity result for partially observed sequences. Proofs of the theoretical results are provided in Supplementary Material.
Recall that at each sequence position $t$ we observe the augmented vector $X_t^\ast$ defined in
\eqref{eq:xtstar-def}. {Under Assumption~\ref{ass:independence}, the null in
\eqref{eqn:nullmissing} makes these observations i.i.d.\ and hence exchangeable. This
derived exchangeability is the basis of the permutation argument, and no particular MCAR or
MAR model is required.}

Let $\mathcal{X}$ denote the sample space of a single augmented observation $X_t^\ast$.
Let $T:\mathcal{X}^n\to\mathbb{R}$ be any deterministic mapping that takes an ordered sequence
$(X_1^\ast,\ldots,X_n^\ast)$ to a scalar test statistic. For any permutation $\pi$ of
$\{1,\dots,n\}$, recall
$
\mathbf{X}^\ast_{\pi} = (X_{\pi(1)}^\ast,\dots,X_{\pi(n)}^\ast)
$
as the permuted sequence, and define the corresponding test statistic
$
T_\pi = T(\mathbf{X}^\ast_{\pi}).
$
Write $T_{\mathrm{obs}}=T(\mathbf{X}^\ast)$ for the statistic computed on the
original sequence, and this is the value corresponding to the identity permutation. Let $S_n$ denote the set of all permutations of $\{1,\dots,n\}$. The full permutation
$p$-value is
\begin{equation}
\label{eq:perm-pvalue}
\hat p_{\mathrm{full}}
= \frac{1}{n!}\sum_{\pi\in S_n} \mathbf{1}\{ T_\pi \ge T_{\mathrm{obs}} \},
\end{equation}

\begin{theorem}
\label{thm:general-permutation}
{Let $T$ be a deterministic statistic as above. Under
Assumption~\ref{ass:independence} and $H_0$ in
\eqref{eqn:nullmissing}, the full permutation $p$-value in
\eqref{eq:perm-pvalue} satisfies
$
{\mathbf P} \bigl( \hat p_{\mathrm{full}} \le \alpha \bigr) \leq \alpha,
 \forall\,\alpha\in[0,1].
$}
\end{theorem}

In practice, exhaustive enumeration over $S_n$ is computationally infeasible. The algorithms in
Section~\ref{Section: framework} therefore use Monte Carlo permutation samples
$\pi_1,\ldots,\pi_B$, drawn independently of the data, and report the implemented approximation
$
\hat p
= \frac{1}{B}\sum_{b=1}^B \mathbf{1}\{T_{\pi_b}\ge T_{\mathrm{obs}}\}.
$
{The full permutation version of \textit{EI-Perm} falls directly under Theorem~\ref{thm:general-permutation} because it recomputes every order-dependent step. For \textit{gMiss}, the additional conditions needed for the fixed \textit{DI} graph are stated in Corollary~\ref{cor:gmiss-permutation}.}

\begin{coro}
\label{cor:gmiss-permutation}
{Under $H_0$ in \eqref{eqn:nullmissing} and
Assumption~\ref{ass:independence}, let
$\hat p_{\mathrm{full}}$ denote the full permutation $p$-value
from a deterministic implementation of either \textit{EI-Perm} or
\textit{gMiss}. For \textit{gMiss}, suppose in addition that the successive MSTs
defining its fixed \textit{DI} graph exist and are almost surely unique. Then
$
{\mathbf P}\bigl(\hat p_{\mathrm{full}}\le\alpha\bigr)
\le\alpha,\alpha\in[0,1].
$
}
\end{coro}

\begin{coro}
\label{cor:pdi-permutation}
{Under $H_0$ in \eqref{eqn:nullmissing} and
Assumption~\ref{ass:independence}, suppose that the successive MSTs
defining the fixed \textit{DI} graph exist and are almost surely unique. Let
$\hat p_{\mathrm{full}}$ denote the full permutation $p$-value of
any deterministic \textit{DI}-based scan statistic computed from the ordered
sequence and this graph. Then
$
{\mathbf P}(\hat p_{\mathrm{full}}\le\alpha)\le\alpha,
\alpha\in[0,1].
$
}
\end{coro}

\section{Numerical studies}
\label{Section: simulation}

\subsection{The choice of \texorpdfstring{{$k$}}{k}}
\label{sec:choiceofk}

Choosing $k$ for similarity graphs such as the $k$-MST remains largely open in change-point detection and related graph-based inference problems \citep{friedman1979multivariate, chen2015graph, chen2019change}. Following recent practice, we parameterize $k=\lfloor n^{\eta}\rfloor$, with $\eta\in(0,1)$ \citep{zhu2024limiting, liu2024generalized, zhou2025asymptotic}.
To assess the sensitivity of \textit{gMiss} to this choice, we conduct a small tuning study over a range of $\eta$ values under representative location and scale alternatives (see Supplementary Material for the full design and results).
Across the considered settings, the power typically increases rapidly as $\eta$ moves away from very small values, and then stabilizes over a moderate range of $\eta$. The empirically best $\eta$ can vary across alternatives, but values around $0.5$--$0.6$ tend to perform robustly.
Accordingly, throughout the numerical studies and real data analysis, we use the $k$-MST with $k=\lfloor n^{0.55}\rfloor$ and Euclidean distance as a practical default rather than a universal optimum.

\subsection{Competing methods and settings}
\label{sec:compare}
We compare \textit{gMiss} with the missing data method \textit{MissInspect} of \citet{follain2022high}, labeled \textit{inspect} in our numerical displays, under light-tailed and heavy-tailed distributions, different covariance structures, and several missingness mechanisms. For the single-change setting, \textit{inspect} provides a canonical scalar score that can be recalculated in a common outer permutation comparison. This outer calibration gives the two procedures a common finite-sample-valid testing framework when all order-dependent steps are recomputed, but it does not alter the alternatives to which each underlying score is sensitive: \textit{MissInspect} is tailored to sparse Gaussian mean shifts, whereas \textit{gMiss} is constructed for changes in the observed data distribution without Gaussianity or sparsity assumptions. The procedure of \citet{liu2025high} outputs an estimated segmentation rather than a canonical scalar test statistic; constructing an outer-permutation test would therefore require an additional choice of global score not specified in that method. \citet{londschien2021change} provide a likelihood-gain statistic, but recalibrating it would require repeated missing-data covariance estimation and graphical-lasso fitting over candidate splits for every permutation. We therefore do not include these procedures in the numerical comparison. Throughout this section, we consider the task of testing for and localizing a single change-point based on independent observations $X_{1}, \ldots, X_{n} \in \R^{p}$.

For each pair $(n, p)$ in $\{200\} \times \{10, 100, 200, 500\}$ and each of the data-generating distributions described below, we use 400 Monte Carlo trials under the null and 200 Monte Carlo trials under each alternative. The public implementation of \textit{MissInspect}\footnote{\url{https://github.com/wangtengyao/MissInspect}} does not return a $p$-value for the projected CUSUM score used in our comparison. We therefore calibrate this score and the \textit{gMiss} score using a common outer permutation procedure. For each simulated dataset, we recompute each method's scalar score on the observed ordering and on 300 permuted orderings, and use the resulting Monte Carlo permutation $p$-value to reject when it is at most $0.05$. Under the null, we report empirical rejection rates. Under the alternatives, we report empirical power and accuracy. Power is the proportion of trials that reject at level $0.05$. Accuracy is the proportion of trials that both reject and have an estimated change-point within 10 sequence positions of the truth. For all tests, the scan maximum is computed over
$[n_0,n_1]$, where $n_0=\lfloor 0.1n\rfloor$ and
$n_1=\lfloor 0.9n\rfloor$. The accuracy calculations use the change-point estimate returned by each method. For \textit{gMiss}, this is the global maximizer of its scan score.

The complete observations are generated according to the following scheme: $X_1,\ldots,X_{\tau}$ are i.i.d. from $F_0$, and $X_{\tau+1}, \ldots, X_n$ are i.i.d. from $F_1$, with $\tau=\lfloor n/3 \rfloor$. In each simulation setting, we select a signal strength that places the methods in a moderate power regime so that performance differences are visible. The location parameters before and after the change-point are denoted by {$\boldsymbol{\mu}_0 \in \R^p$ and $\boldsymbol{\mu}_1 \in \R^p$}, while the scale parameters are denoted by {$\boldsymbol{\Sigma}_0 \in \R^{p \times p}$ and $\boldsymbol{\Sigma}_1 \in \R^{p \times p}$}. Here, a scale alternative refers to a change in the scale parameter $\boldsymbol{\Sigma}_i$. We take {$\boldsymbol{\mu}_0=\mathbf{0}_p$} for $F_0$ and {$\boldsymbol{\mu}_1=\delta\mathbf{1}_p$} for $F_1$, representing a common location shift across coordinates. Under scale alternatives we set $\delta=0$ and let {$\boldsymbol{\Sigma}_1=(1+\sigma)^2\boldsymbol{\Sigma}_0$}. Specifically, we consider the following distributions for $F_i$, with $i=0,1$:

(I) the multivariate Gaussian distribution {$N_p(\boldsymbol{\mu}_i, \boldsymbol{\Sigma}_i)$}, {$\boldsymbol{\Sigma}_0=\mathbf{I}_p$};

(II) the multivariate Gaussian distribution {$N_p(\boldsymbol{\mu}_i, \boldsymbol{\Sigma}_i)$}, {$(\boldsymbol{\Sigma}_0)_{kj}=0.6^{|k-j|}$}, $k,j \in [p]$;

(III) the multivariate $t_3$ distribution {$t_3(\boldsymbol{\mu}_i, \boldsymbol{\Sigma}_i)$}, {$\boldsymbol{\Sigma}_0=\mathbf{I}_p$};

(IV) the multivariate lognormal distribution {$\exp\{N_p(\boldsymbol{\mu}_i, \boldsymbol{\Sigma}_i)\}$, where exponentiation is applied coordinatewise}, with {$\boldsymbol{\Sigma}_0=\mathbf{I}_p$}.\\
For each setting, we consider two alternatives:

(a) location ({$\delta > 0$ and $\boldsymbol{\Sigma}_1 = \boldsymbol{\Sigma}_0$}),

(b) scale ({$\delta=0$ and $\boldsymbol{\Sigma}_1= (1+\sigma)^2 \boldsymbol{\Sigma}_0$}).\\
The parameters $\delta$ and $\sigma$ {vary by setting}, as shown in Table~\ref{tab:simu,signal}. We consider two MCAR mechanisms as well as an additional MAR mechanism, which is often more realistic in applications where missingness depends on observed values. For $i \in [p], t \in [n]$,
\par\noindent(1) \textit{Homogeneous MCAR.} $M_{it}\stackrel{\mathrm{i.i.d.}}{\sim}\mathrm{Bernoulli}(q)$ with
$q\in\{0.5,0.6,\ldots,1\}$.

\noindent(2) \textit{Heterogeneous MCAR.} For each dimension $p$, draw $q_i\overset{\mathrm{i.i.d.}}{\sim}\mathrm{Uniform}(0.5,0.9)$ for $i=1,\ldots,p$ and hold the resulting vector fixed across Monte Carlo trials. Conditional on $\{q_i\}_{i=1}^p$, generate $M_{it}\overset{\mathrm{ind}}{\sim}\mathrm{Bernoulli}(q_i)$ independently over $i$ and $t$.

\noindent(3) \textit{MAR based on an anchor.}
Let $a=\lfloor p/4\rfloor$ and $A=\{1,\ldots,a\}$ be an anchor subset of coordinates that are always observed, i.e., $M_{it}=1$ for all $i\in A$ and all $t$. Define the observed summary
$
\check{X}_t=\frac{1}{a}\sum_{i\in A}X_{it}.
$
For each coordinate $i\in [p]\setminus A$, we generate
$M_{it}\mid \check{X}_t \sim \mathrm{Bernoulli}\bigl(\mathrm{expit}(\beta+\check{X}_t)\bigr)$,
where $\mathrm{expit}(x)=\frac{1}{1+e^{-x}}$,
independently over $(i,t)$ conditional on $\{\check{X}_t\}_{t=1}^n$, and $\beta\in\{0.5,0.6,\ldots,1\}$.

{This MAR construction produces augmented observations that are independent across sequence positions and therefore satisfies Assumption~\ref{ass:independence}. Under $H_0$, applying the same rule at every sequence position also gives a common marginal distribution, so $X_1^\ast,\ldots,X_n^\ast$ are i.i.d.\ and full permutation calibration applies.} Results are reported in Supplementary Material.

\begin{table}[!htbp]
\caption{Signal strengths for the location and scale alternatives.}
\label{tab:simu,signal}
\centering
\footnotesize
\renewcommand{\arraystretch}{1.35}
\begin{tabular}{lcccc}
\toprule
\textbf{Setting} & (I) & (II) & (III) & (IV) \\
\midrule
(a): $\delta$
& $\dfrac{2\log(p)}{5p^{0.5}}$
& $\dfrac{2\log(p)}{5p^{0.5}}$
& $\dfrac{7\log(p)}{20p^{0.47}}$
& $\dfrac{3\log(p)}{4p^{0.65}}$
\\[6pt]
(b): $\sigma$
& $\dfrac{7\log(p)}{17p^{0.8}}$
& $\dfrac{2\log(p)}{5p^{0.7}}$
& $\dfrac{3\log(p)}{10p^{0.33}}$
& $\dfrac{7\log(p)}{10p^{0.75}}$
\\
\bottomrule
\end{tabular}
\end{table}

The power and accuracy under homogeneous missingness are displayed in Figures~\ref{fig:power,missrate} and~\ref{fig:accuracy,missrate}, where the left panels correspond to location alternatives and the right panels to scale alternatives. Under Gaussian location alternatives, \textit{gMiss} and \textit{inspect} have comparable power across dimensions, with \textit{inspect} often slightly higher. For moderate and high dimensions, \textit{gMiss} more often both rejects and localizes the change within the specified window.

For the non-Gaussian location alternatives, especially multivariate $t_3$ and lognormal data, \textit{gMiss} has higher power and accuracy in most plotted settings. Under scale alternatives, \textit{gMiss} maintains nontrivial performance on both measures across the four distributional settings, whereas \textit{inspect} has low power and accuracy near zero in these experiments. These patterns are consistent with the design of \textit{inspect}, which primarily targets Gaussian location shifts, and with the broader empirical scope of \textit{gMiss} in the simulation designs considered here.

\begin{figure}[!htbp]
\centering
\begin{minipage}{1\textwidth}
  \centering
\includegraphics[width=1\linewidth]{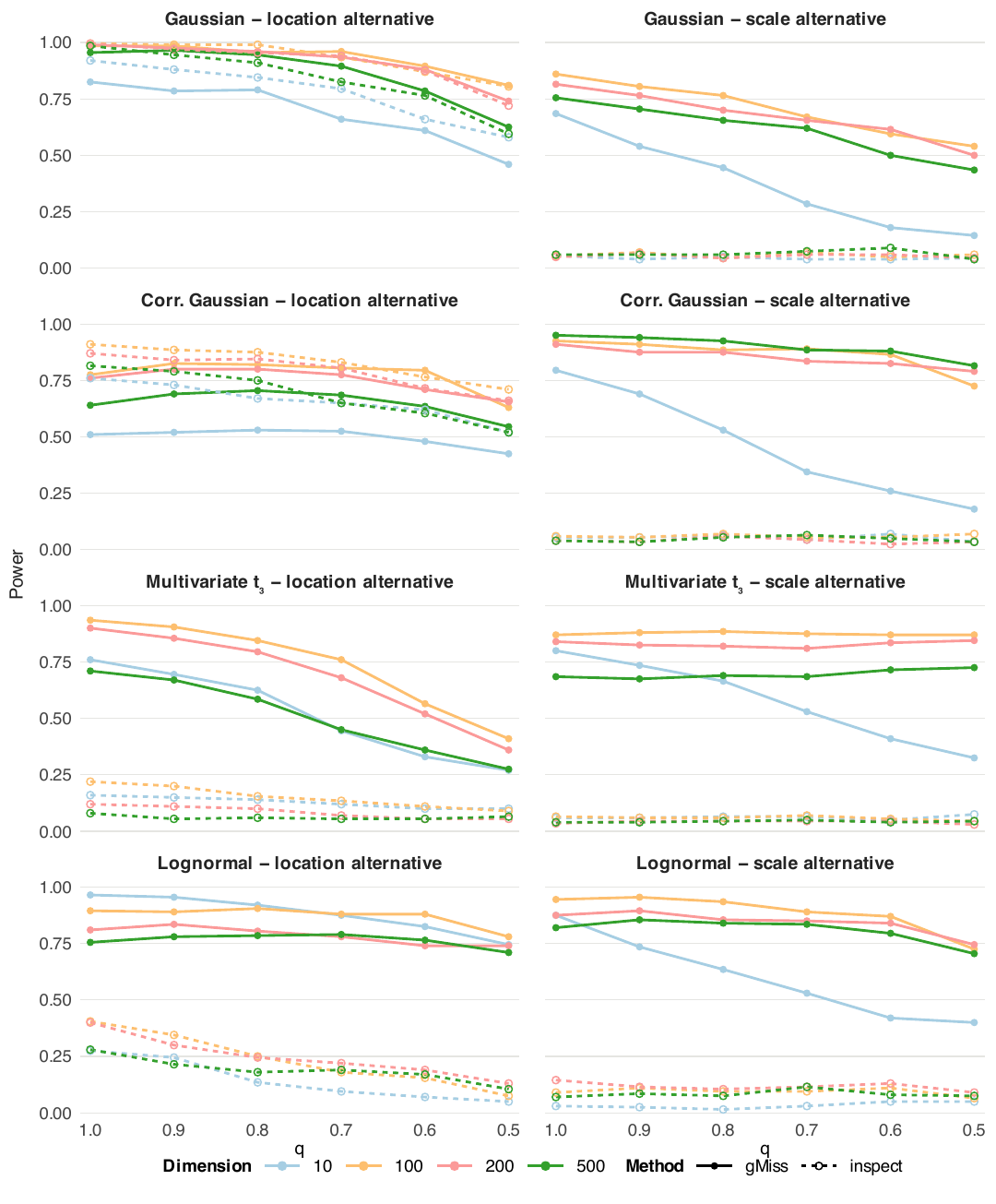}
\caption{Power comparison of \textit{gMiss} and \textit{inspect} under homogeneous MCAR missingness as a function of the observation rate $q$ (missing rate $1-q$).}
\label{fig:power,missrate}
\end{minipage}
\end{figure}

\begin{figure}[!htbp]
\centering
\begin{minipage}{1\textwidth}
  \centering
\includegraphics[width=1\linewidth]{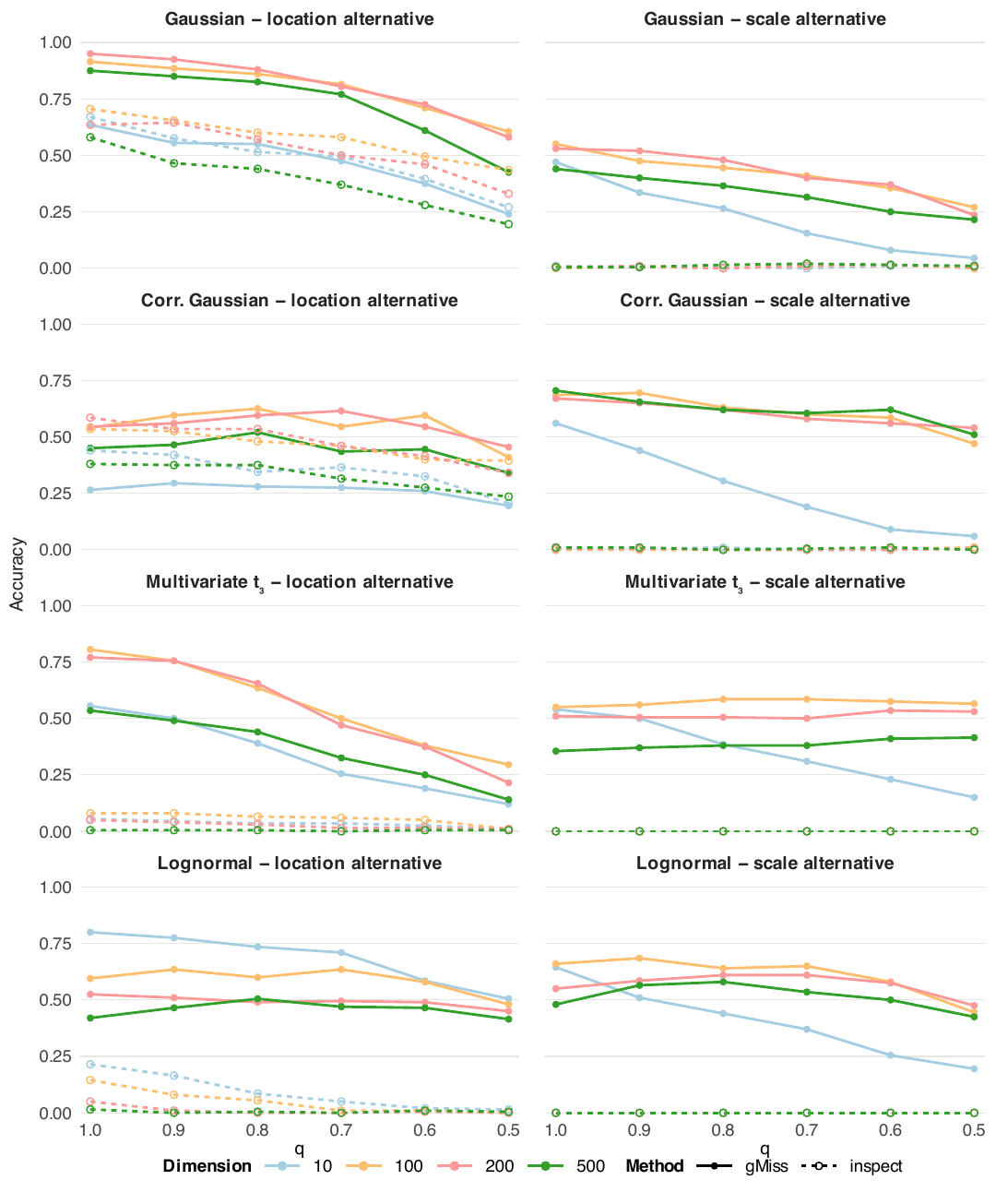}
\caption{Accuracy of \textit{gMiss} and \textit{inspect} under homogeneous MCAR missingness as a function of the observation rate $q$ (missing rate $1-q$).}
\label{fig:accuracy,missrate}
\end{minipage}
\end{figure}

\begin{table}
\centering
\caption{Power and accuracy (in parentheses) for \textit{gMiss} and \textit{inspect} under heterogeneous MCAR ($\alpha=0.05$). Boldface marks the larger point estimate within each method pair.}
\label{T:hetermiss}
\begin{adjustbox}{width=0.9\textwidth,center}
\begin{tabular}{c | c |c c |cc|cc|cc}
\hline\hline
\multirow{2}{*}{} & \multirow{2}{*}{\text{Distribution}} & \multicolumn{2}{|c|}{\( p = 10 \)} & \multicolumn{2}{c|}{\( p = 100 \)} & \multicolumn{2}{c}{\( p = 200 \)} & \multicolumn{2}{c}{\( p = 500 \)}\\ \cline{3-10}
&  & \textit{gMiss} &  \textit{inspect}  & \textit{gMiss} &  \textit{inspect}  & \textit{gMiss} &  \textit{inspect}   & \textit{gMiss} &  \textit{inspect} \\ \hline
\multirow{8}{*}{\shortstack{{location}\\{alternative}}}
& \multirow{2}{*}{Gaussian}
  & 0.690 & \textbf{0.770} & 0.905 & \textbf{0.920} & \textbf{0.960} & 0.930 & 0.855 & \textbf{0.865}\\
&
  & (0.490) & (\textbf{0.505}) & (\textbf{0.765}) & (0.525) & (\textbf{0.805}) & (0.595) & (\textbf{0.680}) & (0.380) \\ \cline{2-10}
& \multirow{2}{*}{Corr.\ Gaussian}
  & 0.490 & \textbf{0.680} & 0.730 & \textbf{0.815} & \textbf{0.770} & 0.765 & 0.690 & \textbf{0.710}\\
&
  & (0.255) & (\textbf{0.315}) & (\textbf{0.500}) & (0.495) & (\textbf{0.500}) & (0.415) & (\textbf{0.460}) & (0.345)\\ \cline{2-10}
& \multirow{2}{*}{multivariate $t_3$}
  & \textbf{0.410} & 0.125 & \textbf{0.680} & 0.145 & \textbf{0.615} & 0.105 & \textbf{0.375} & 0.070\\
&
  & (\textbf{0.235}) & (0.055) & (\textbf{0.505}) & (0.030) & (\textbf{0.420}) & (0.025) & (\textbf{0.305}) & (0.010)\\ \cline{2-10}
& \multirow{2}{*}{lognormal}
  & \textbf{0.860} & 0.115 & \textbf{0.875} & 0.130 & \textbf{0.800} & 0.195 & \textbf{0.770} & 0.180\\
&
  & (\textbf{0.685}) & (0.090) & (\textbf{0.600}) & (0.010) & (\textbf{0.480}) & (0.000) & (\textbf{0.460}) & (0.000)\\
\hline
\multirow{8}{*}{\shortstack{{scale}\\{alternative}}}
  & \multirow{2}{*}{Gaussian}
  & \textbf{0.355} & 0.035 & \textbf{0.650} & 0.060 & \textbf{0.680} & 0.045 & \textbf{0.620} & 0.060\\
  &
    & (\textbf{0.165}) & (0.000) & (\textbf{0.320}) & (0.010) & (\textbf{0.370}) & (0.010) & (\textbf{0.300}) & (0.015) \\
\cline{2-10}
  & \multirow{2}{*}{Corr.\ Gaussian}
  & \textbf{0.435} & 0.045 & \textbf{0.870} & 0.065 & \textbf{0.865} & 0.080 & \textbf{0.895} & 0.060\\
  &
    & (\textbf{0.250}) & (0.005) & (\textbf{0.575}) & (0.005) & (\textbf{0.605}) & (0.010) & (\textbf{0.575}) & (0.005)\\
\cline{2-10}
  & \multirow{2}{*}{multivariate $t_3$}
  & \textbf{0.555} & 0.060 & \textbf{0.850} & 0.065 & \textbf{0.810} & 0.045 & \textbf{0.690} & 0.040\\
  &
    & (\textbf{0.330}) & (0.000) & (\textbf{0.520}) & (0.000) & (\textbf{0.475}) & (0.000) & (\textbf{0.375}) & (0.000)\\
\cline{2-10}
  & \multirow{2}{*}{lognormal}
  & \textbf{0.540} & 0.030 & \textbf{0.900} & 0.060 & \textbf{0.850} & 0.110 & \textbf{0.830} & 0.100\\
  &
    & (\textbf{0.300}) & (0.000) & (\textbf{0.590}) & (0.000) & (\textbf{0.550}) & (0.000) & (\textbf{0.505}) & (0.000)\\
\hline\hline
\end{tabular}
\end{adjustbox}
\end{table}

Table~\ref{T:hetermiss} gives the corresponding results under heterogeneous MCAR. For Gaussian location alternatives, \textit{inspect} has higher power at $p=10$, while the gap narrows as the dimension increases. For multivariate $t_3$ and lognormal location alternatives, \textit{gMiss} has higher power and accuracy across the reported dimensions. Under scale alternatives, \textit{inspect} has low values on both measures, whereas \textit{gMiss} retains nontrivial performance.

\section{A real data application}
\label{Section: real data}

Array Comparative Genomic Hybridization (aCGH) is a microarray technique for measuring DNA copy-number changes along the genome. In tumor studies, abrupt copy-number changes correspond to genomic breakpoints, so multiple change-point detection provides a natural tool for identifying candidate copy-number variation regions. The bladder tumor dataset of \citet{stransky2006regional} therefore offers a useful test bed for our method: it is high-dimensional, partially observed, and expected to contain multiple local genomic alterations.

The dataset contains 57 bladder tumor profiles, each observed over 2385 ordered probe locations, with approximately 7\% missing entries in total. Earlier analyses of this dataset include the group fused lasso approach of \citet{Bleakley2011TheGF} and the \textit{ecp} analysis of \citet{matteson2014nonparametric}. Notably, \citet{matteson2014nonparametric} removed series with substantial missingness and analyzed only 43 individuals. In contrast, our framework allows us to retain and analyze all 57 profiles directly. To facilitate comparison with this earlier analysis, we also report \textit{ecp} below.

We compare \textit{gMiss} with \textit{inspect} \citep{follain2022high} and \textit{ecp}.
For \textit{gMiss}, we use the construction above without robust standardization and obtain
multiple candidates through a Wild Binary Segmentation-type recursion, following the
graph-based candidate-search strategy of \citet{zhang2021graph}. For \textit{inspect}, we use
a binary segmentation wrapper around the authors' single change-point implementation, retaining
its built-in standardization. For \textit{ecp}, we perform elementwise imputation once and
then apply the energy-based divisive procedure implemented by \texttt{e.divisive} in the
\texttt{ecp} R package. The \textit{gMiss} and \textit{inspect} searches use 3000 outer permutations at each
tested interval or segment, while \textit{ecp} uses 3000 internal permutations. Because this calibration follows one-time imputation rather than reproducing the imputation step within each permutation, it can be size-invalid, as illustrated in Table~\ref{tab:moti2}. An outer recalibration is possible in principle, but the \textit{ecp} results below use the internal calibration and are interpreted descriptively. For candidate
generation, the recursive searches use a local threshold of $0.05$ to decide whether to
continue splitting. We subsequently display candidates with recorded $p$-values at most
$0.00034$, a stringent cutoff near the Monte Carlo resolution induced by 3000 permutations. The
resulting change-points are treated as exploratory candidates rather than formal discoveries.

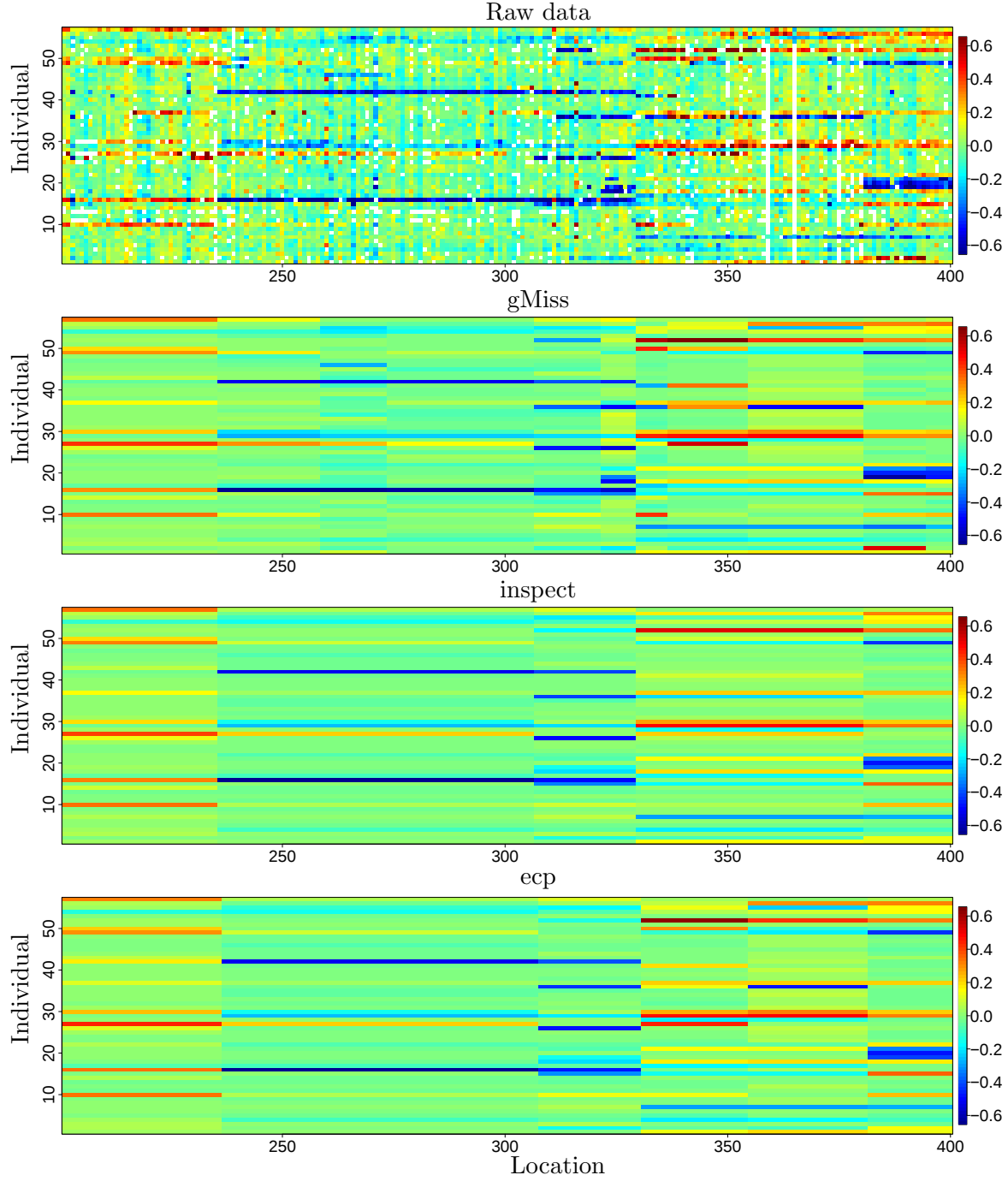
\begin{figure}[!b]
\centering

\begin{minipage}{0.98\linewidth}
\centering
\begin{overpic}[width=0.95\linewidth]{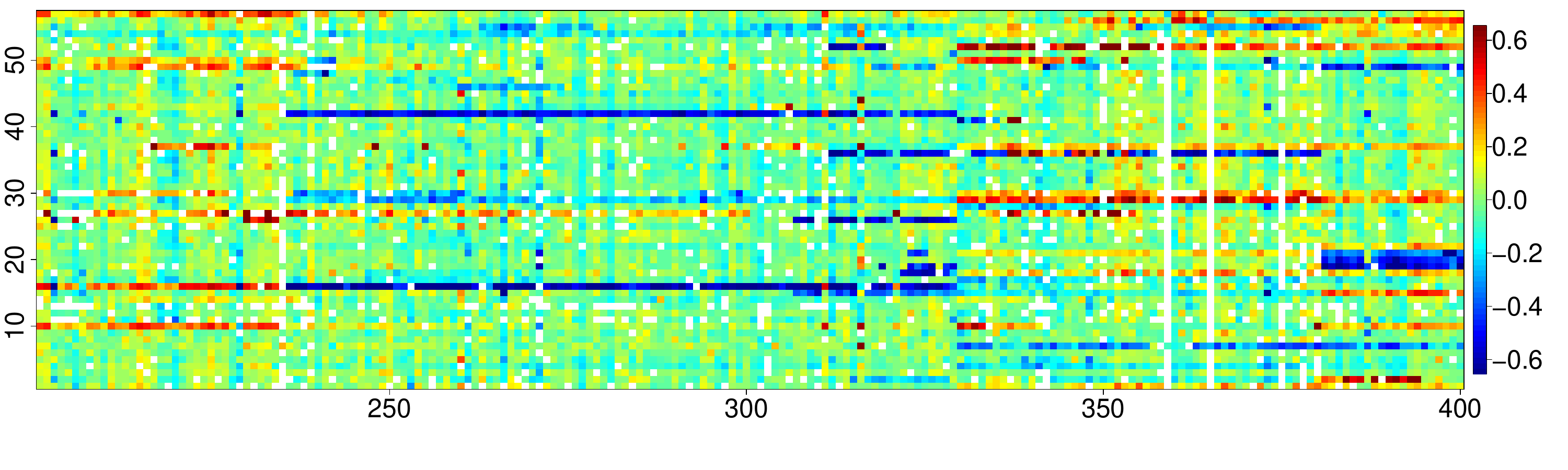}

\put(51,31){\makebox(0,0){\small Raw data}}
\put(-2,20){\rotatebox{90}{\makebox(0,0){\small Individual}}}
\end{overpic}
\end{minipage}

\vspace{0.15em}

\begin{minipage}{0.98\linewidth}
\centering
\begin{overpic}[width=0.95\linewidth]{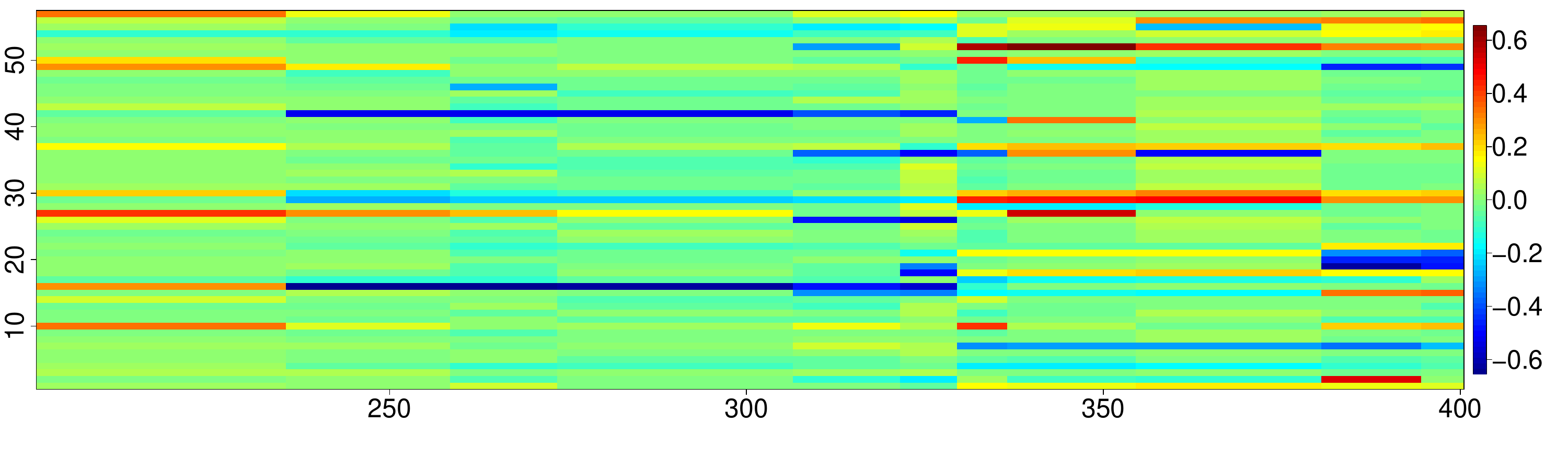}

\put(51,31){\makebox(0,0){\small gMiss}}
\put(-2,20){\rotatebox{90}{\makebox(0,0){\small Individual}}}
\end{overpic}

\end{minipage}

\vspace{0.15em}

\begin{minipage}{0.98\linewidth}
\centering
\begin{overpic}[width=0.95\linewidth]{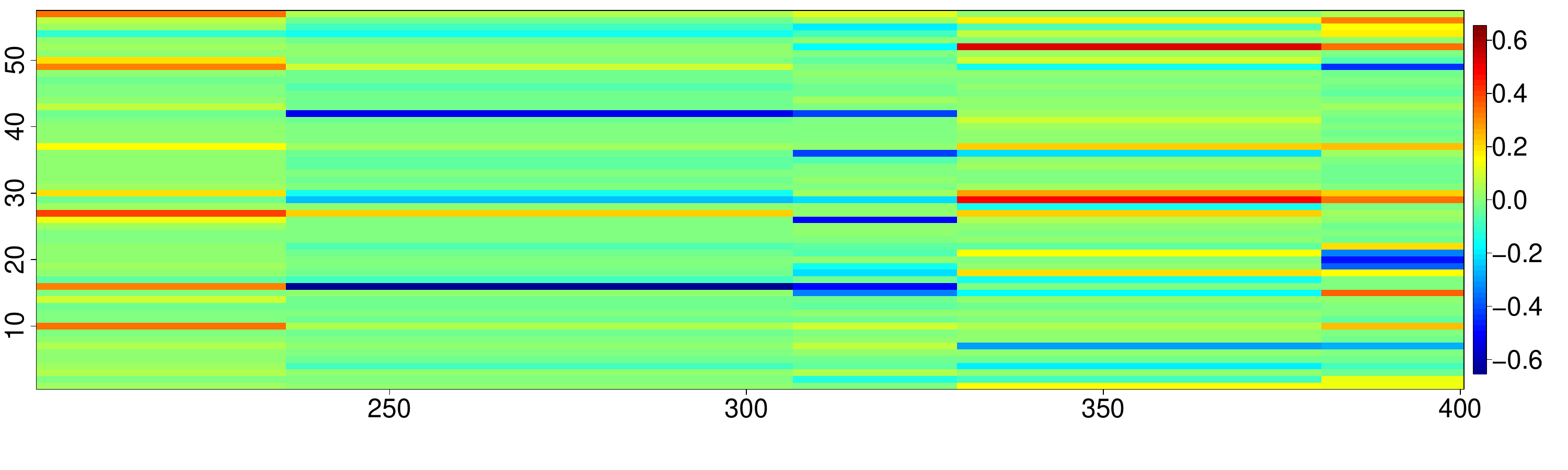}
\put(51,31){\makebox(0,0){\small inspect}}
\put(-2,20){\rotatebox{90}{\makebox(0,0){\small Individual}}}
\end{overpic}

\end{minipage}

\vspace{0.15em}

\begin{minipage}{0.98\linewidth}
\centering
\begin{overpic}[width=0.95\linewidth]{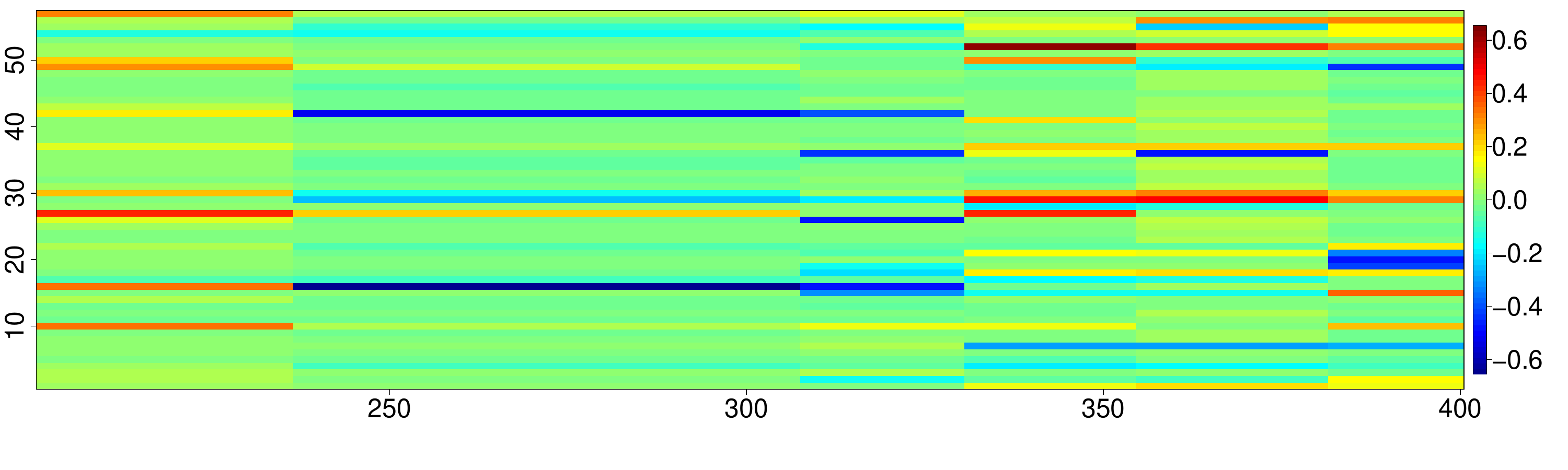}

\put(51,31){\makebox(0,0){\small ecp}}
\put(-2,20){\rotatebox{90}{\makebox(0,0){\small Individual}}}
\put(53,2){\makebox(0,0){\small Location}}
\end{overpic}

\end{minipage}

\caption{Heatmaps for probe locations 201--400 under the raw data and three competing methods. White cells in the raw data panel indicate missing entries.}
\label{fig:realdata_201_400_all}
\end{figure}

{Applying this screening rule to the full dataset yields 139 candidate change-points for \textit{gMiss}, compared with 79 for \textit{inspect} and 72 for \textit{ecp}, while the version of \textit{gMiss} with robust standardization yields 134. We do not regard the larger number alone
as evidence of superiority. Rather, our main point is that many of the additional candidate
boundaries highlighted by \textit{gMiss} appear visually plausible in the raw data heatmap. The main text examines probe locations 201--400, while Supplementary Material reports the remaining regions and the results with robust standardization. For visualization, the top panel shows the raw data over the selected probe range. The lower panels show a piecewise constant reconstruction for each method, obtained by replacing the observations between adjacent estimated change-points by their within-segment sample mean for each profile.}

In Figure~\ref{fig:realdata_201_400_all}, \textit{gMiss} identifies additional changes near probe locations 320 and 395 that are not identified by either \textit{inspect} or \textit{ecp}. To assess whether these differences are driven by the stringent display cutoff, we also examine the results using a cutoff of 0.01. At this more lenient cutoff, neither comparator identifies the change near 320, while \textit{inspect}, but not \textit{ecp}, identifies the change near 395. The corresponding heatmaps are shown in Supplementary Material.

\section{Conclusion}
\label{Sec: discussion}

\textit{gMiss} provides a nonparametric framework for testing and localizing a change in the observed data distribution of a partially observed high-dimensional sequence, without Gaussianity or sparsity assumptions. It combines an \textit{EI}-based weighted scan and a \textit{DI}-based difference scan to capture complementary graph patterns. A central feature of the full-permutation version is finite-sample type I error control. The simulations show competitive performance under Gaussian location alternatives and gains in many non-Gaussian location and scale settings, while the aCGH analysis shows that it can identify additional visually plausible boundaries in partially observed high-dimensional data.

\section*{Acknowledgments}

This work was supported in part by the U.S. National Science Foundation under grants DMS-1848579 and DMS-2311399.

\bibliographystyle{apalike}
{
\setlength{\bibsep}{0pt}
\bibliography{library}

@article{matteson2014nonparametric,
  title={A nonparametric approach for multiple change point analysis of multivariate data},
  author={Matteson, David S and James, Nicholas A},
  journal={Journal of the American Statistical Association},
  volume={109},
  number={505},
  pages={334--345},
  year={2014},
  publisher={Taylor \& Francis}
}

@article{xie2012change,
  title={Change-point detection for high-dimensional time series with missing data},
  author={Xie, Yao and Huang, Jiaji and Willett, Rebecca},
  journal={IEEE Journal of Selected Topics in Signal Processing},
  volume={7},
  number={1},
  pages={12--27},
  year={2013},
  publisher={IEEE}
}

@article{londschien2021change,
  title={Change-point detection for graphical models in the presence of missing values},
  author={Londschien, Malte and Kov{\'a}cs, Solt and B{\"u}hlmann, Peter},
  journal={Journal of Computational and Graphical Statistics},
  volume={30},
  number={3},
  pages={768--779},
  year={2021},
  publisher={Taylor \& Francis}
}

@article{chu2019asymptotic,
author = {Lynna Chu and Hao Chen},
title = {{Asymptotic distribution-free change-point detection for multivariate and non-Euclidean data}},
volume = {47},
journal = {The Annals of Statistics},
number = {1},
publisher = {Institute of Mathematical Statistics},
pages = {382 -- 414},
year = {2019},
doi = {10.1214/18-AOS1691},
URL = {https://doi.org/10.1214/18-AOS1691}
}

@article{liu2022fast,
  title={A Fast and Efficient Change-Point Detection Framework Based on Approximate $ k $-Nearest Neighbor Graphs},
  author={Liu, Yi-Wei and Chen, Hao},
  journal={IEEE Transactions on Signal Processing},
  volume={70},
  pages={1976--1986},
  year={2022},
  publisher={IEEE}
}

@inproceedings{harchaoui2007retrospective,
  title={Retrospective mutiple change-point estimation with kernels},
  author={Harchaoui, Zaid and Capp{\'e}, Olivier},
  booktitle={2007 IEEE/SP 14th Workshop on Statistical Signal Processing},
  pages={768--772},
  year={2007},
  organization={IEEE}
}

@article{harchaoui2008kernel,
  title={Kernel change-point analysis},
  author={Harchaoui, Zaid and Moulines, Eric and Bach, Francis},
  journal={Advances in neural information processing systems},
  volume={21},
  year={2008}
}

@article{li2015m,
  title={M-statistic for kernel change-point detection},
  author={Li, Shuang and Xie, Yao and Dai, Hanjun and Song, Le},
  journal={Advances in Neural Information Processing Systems},
  volume={28},
  year={2015}
}

@article{zhang2021graph,
  title={Graph-based multiple change-point detection},
  author={Zhang, Yuxuan and Chen, Hao},
  journal={arXiv preprint arXiv:2110.01170},
  year={2021}
}

@article{Bleakley2011TheGF,
  title={The group fused Lasso for multiple change-point detection},
  author={Kevin Bleakley and Jean-Philippe Vert},
  journal={arXiv: Quantitative Methods},
  year={2011}
}

@article{stransky2006regional,
  title={Regional copy number--independent deregulation of transcription in cancer},
  author={Stransky, Nicolas and Vallot, C{\'e}line and Reyal, Fabien and Bernard-Pierrot, Isabelle and De Medina, Sixtina Gil Diez and Segraves, Rick and De Rycke, Yann and Elvin, Paul and Cassidy, Andrew and Spraggon, Carolyn and others},
  journal={Nature genetics},
  volume={38},
  number={12},
  pages={1386--1396},
  year={2006},
  publisher={Nature Publishing Group US New York}
}

@article{arlot2019kernel,
  title={A kernel multiple change-point algorithm via model selection},
  author={Arlot, Sylvain and Celisse, Alain and Harchaoui, Zaid},
  journal={Journal of machine learning research},
  volume={20},
  number={162},
  year={2019}
}

@article{chen2019universal,
  title={A universal nonparametric event detection framework for neuropixels data},
  author={Chen, Hao and Chen, Shizhe and Deng, Xinyi},
  journal={bioRxiv},
  pages={650671},
  year={2019},
  publisher={Cold Spring Harbor Laboratory}
}

@article{friedman1979multivariate,
  title={Multivariate generalizations of the Wald-Wolfowitz and Smirnov two-sample tests},
  author={Friedman, Jerome H and Rafsky, Lawrence C},
  journal={The Annals of Statistics},
  pages={697--717},
  year={1979},
  publisher={JSTOR}
}

@article{zhu2024limiting,
  title={Limiting distributions of graph-based test statistics on sparse and dense graphs},
  author={Zhu, Yejiong and Chen, Hao},
  journal={Bernoulli},
  volume={30},
  number={1},
  pages={770--796},
  year={2024},
  publisher={Bernoulli Society for Mathematical Statistics and Probability}
}

@article{chen2019change,
  title={Change-point detection for multivariate and non-euclidean data with local dependency},
  author={Chen, Hao},
  journal={arXiv preprint arXiv:1903.01598},
  year={2019}
}

@article{song2024practical,
  author={Song, Hoseung and Chen, Hao},
  journal={IEEE Transactions on Signal Processing}, 
  title={Practical and Powerful Kernel-Based Change-Point Detection}, 
  year={2024},
  volume={72},
  number={},
  pages={5174-5186},
  doi={10.1109/TSP.2024.3479274}}

@article{jiang2023robust,
  title={Robust inference for change points in high dimension},
  author={Jiang, Feiyu and Wang, Runmin and Shao, Xiaofeng},
  journal={Journal of Multivariate Analysis},
  volume={193},
  pages={105114},
  year={2023},
  publisher={Elsevier}
}

@article{loh2011high,
  title={High-dimensional regression with noisy and missing data: Provable guarantees with non-convexity},
  author={Loh, Po-Ling and Wainwright, Martin J},
  journal={Advances in neural information processing systems},
  volume={24},
  year={2011}
}

@article{troyanskaya2001missing,
  title={Missing value estimation methods for DNA microarrays},
  author={Troyanskaya, Olga and Cantor, Michael and Sherlock, Gavin and Brown, Pat and Hastie, Trevor and Tibshirani, Robert and Botstein, David and Altman, Russ B},
  journal={Bioinformatics},
  volume={17},
  number={6},
  pages={520--525},
  year={2001},
  publisher={Oxford University Press}
}

@article{follain2022high,
  title={High-dimensional changepoint estimation with heterogeneous missingness},
  author={Follain, Bertille and Wang, Tengyao and Samworth, Richard J},
  journal={Journal of the Royal Statistical Society Series B: Statistical Methodology},
  volume={84},
  number={3},
  pages={1023--1055},
  year={2022},
  publisher={Oxford University Press}
}

@article{rosenbaum2005exact,
  title={An exact distribution-free test comparing two multivariate distributions based on adjacency},
  author={Rosenbaum, Paul R},
  journal={Journal of the Royal Statistical Society Series B: Statistical Methodology},
  volume={67},
  number={4},
  pages={515--530},
  year={2005},
  publisher={Oxford University Press}
}

@article{liu2024generalized,
  title={Generalized Independence Test for Modern Data},
  author={Liu, Mingshuo and Zhou, Doudou and Chen, Hao},
  journal={arXiv preprint arXiv:2409.07745},
  year={2024}
}

@article{liu2025high,
  title={High-dimensional change point detection with missing values},
  author={Liu, Yanxi and Safikhani, Abolfazl},
  journal={Electronic Journal of Statistics},
  volume={19},
  number={2},
  pages={5019--5067},
  year={2025},
  publisher={The Institute of Mathematical Statistics and the Bernoulli Society}
}

@article{barnett2016change,
  title={Change point detection in correlation networks},
  author={Barnett, Ian and Onnela, Jukka-Pekka},
  journal={Scientific reports},
  volume={6},
  number={1},
  pages={18893},
  year={2016},
  publisher={Nature Publishing Group UK London}
}

@article{andreou2009structural,
  title={Structural breaks in financial time series},
  author={Andreou, Elena and Ghysels, Eric},
  journal={Handbook of financial time series},
  pages={839--870},
  year={2009},
  publisher={Springer}
}

@article{banerjee2020change,
  title={Change-point analysis in financial networks},
  author={Banerjee, Sayantan and Guhathakurta, Kousik},
  journal={Stat},
  volume={9},
  number={1},
  pages={e269},
  year={2020},
  publisher={Wiley Online Library}
}

@article{allen2018non,
  title={Non-parametric multiple change point analysis of the global financial crisis},
  author={Allen, David E and McAleer, Michael and Powell, Robert J and Singh, Abhay K},
  journal={Annals of Financial Economics},
  volume={13},
  number={02},
  pages={1850008},
  year={2018},
  publisher={World Scientific}
}

@article{wang2021optimal,
  title={Optimal change point detection and localization in sparse dynamic networks},
  author={Wang, Daren and Yu, Yi and Rinaldo, Alessandro},
  journal={The Annals of Statistics},
  volume={49},
  number={1},
  pages={203--232},
  year={2021},
  publisher={JSTOR}
}

@inproceedings{peel2015detecting,
  title={Detecting change points in the large-scale structure of evolving networks},
  author={Peel, Leto and Clauset, Aaron},
  booktitle={Proceedings of the AAAI conference on artificial intelligence},
  volume={29},
  number={1},
  year={2015}
}

@article{reeves2007review,
  title={A review and comparison of changepoint detection techniques for climate data},
  author={Reeves, Jaxk and Chen, Jien and Wang, Xiaolan L and Lund, Robert and Lu, Qi Qi},
  journal={Journal of applied meteorology and climatology},
  volume={46},
  number={6},
  pages={900--915},
  year={2007}
}

@article{shi2022changepoint,
  title={Changepoint detection: An analysis of the Central England temperature series},
  author={Shi, Xueheng and Beaulieu, Claudie and Killick, Rebecca and Lund, Robert},
  journal={Journal of Climate},
  volume={35},
  number={19},
  pages={6329--6342},
  year={2022}
}

@article{garreau2018consistent,
author = {Damien Garreau and Sylvain Arlot},
title = {{Consistent change-point detection with kernels}},
volume = {12},
journal = {Electronic Journal of Statistics},
number = {2},
publisher = {Institute of Mathematical Statistics and Bernoulli Society},
pages = {4440 -- 4486},
year = {2018},
doi = {10.1214/18-EJS1513},
URL = {https://doi.org/10.1214/18-EJS1513}
}

@article{zhang2022adaptive,
  title={Adaptive inference for change points in high-dimensional data},
  author={Zhang, Yangfan and Wang, Runmin and Shao, Xiaofeng},
  journal={Journal of the American Statistical Association},
  volume={117},
  number={540},
  pages={1751--1762},
  year={2022},
  publisher={Taylor \& Francis}
}

@article{li2020asymptotic,
  title={Asymptotic distribution-free change-point detection based on interpoint distances for high-dimensional data},
  author={Li, Jun},
  journal={Journal of Nonparametric Statistics},
  volume={32},
  number={1},
  pages={157--184},
  year={2020},
  publisher={Taylor \& Francis}
}

@article{wang2022inference,
  title={Inference for change points in high-dimensional data via selfnormalization},
  author={Wang, Runmin and Zhu, Changbo and Volgushev, Stanislav and Shao, Xiaofeng},
  journal={The Annals of Statistics},
  volume={50},
  number={2},
  pages={781--806},
  year={2022},
  publisher={Institute of Mathematical Statistics}
}

@article{wells2013strategies,
  title={Strategies for handling missing data in electronic health record derived data},
  author={Wells, Brian J and Chagin, Kevin M and Nowacki, Amy S and Kattan, Michael W},
  journal={Egems},
  volume={1},
  number={3},
  pages={1035},
  year={2013}
}

@article{zhou2025asymptotic,
  author={Zhou, Doudou and Chen, Hao},
  journal={IEEE Transactions on Information Theory}, 
  title={Asymptotic Distribution-Free Change-Point Detection for Modern Data Based on a New Ranking Scheme}, 
  year={2025},
  volume={71},
  number={8},
  pages={6183-6197},
  doi={10.1109/TIT.2025.3575858}}

@article{chen2015graph,
author = {Hao Chen and Nancy Zhang},
title = {{Graph-based change-point detection}},
volume = {43},
journal = {The Annals of Statistics},
number = {1},
publisher = {Institute of Mathematical Statistics},
pages = {139 -- 176},
year = {2015},
doi = {10.1214/14-AOS1269},
URL = {https://doi.org/10.1214/14-AOS1269}
}
}
\end{document}